\documentclass[12pt]{iopart}
\usepackage{graphicx}
\renewcommand{\a}{\displaystyle}

\usepackage{dcolumn}
\usepackage{bm}
\usepackage{verbatim}
\usepackage{setspace}
\usepackage{ulem}

\begin{document}

\title{Toward an Ising Model of Cancer and Beyond}

Salvatore Torquato$^{1,2,3,4,5}$

$^1$  Department of Chemistry,
Princeton University,
Princeton, NJ 08544, USA

$^2$ Department of Physics,
Princeton University,
Princeton, NJ 08544, USA

$^3$ Princeton Center for Theoretical Science,
Princeton University,
Princeton, NJ 08544, USA

$^4$ Program in Applied and Computational Mathematics,
Princeton University,
Princeton, NJ 08544, USA

$^5$ Princeton Institute for the Science and Technology of Materials,
Princeton University,
Princeton, NJ 08544, USA

\vspace{1in} \noindent \textbf{Corresponding author contact information}:
\begin{tabbing}
 \hspace{0.25in}
 \= Salvatore Torquato \\
 \>  Tel.: 609-258-3341 \\
 \>  Fax: 609-258-6746 \\
 \>  E-mail: torquato$@$princeton.edu
\end{tabbing}

\noindent \textbf{Short title}: Toward an Ising Model of Cancer \newline

\noindent \textbf{Classification numbers}: 87.17.Aa, 87.19.xj

\newpage

\begin{abstract}

The holy grail of  tumor modeling is to formulate theoretical and computational
tools that can be utilized in the clinic to predict neoplastic progression
and propose individualized optimal treatment strategies to control cancer
growth. In order to develop such a predictive model, one must account for
the numerous complex mechanisms involved in tumor growth. Here we review resarch  work that we 
have  done toward the development of an``Ising model'' of cancer.  The Ising model
is an idealized statistical-mechanical model of ferromagnetism
that is based on simple local-interaction rules, but nonetheless
leads to basic insights and features of real magnets, such as phase transitions with a critical point.
The review  begins with a description of a minimalist four-dimensional (three dimensions
in space and one in time) cellular automaton (CA) model
of cancer in which healthy cells transition between states (proliferative, hypoxic, and necrotic)
according to simple local rules and their present states, which
can viewed as a stripped-down Ising model of cancer. This model
is applied to model the growth of glioblastoma multiforme, the most malignant of brain cancers.
This is followed by a discussion of the extension of the model
to study the effect on the tumor
dynamics and geometry of a mutated subpopulation.
A discussion of how tumor growth is affected by chemotherapeutic
treatment, including induced resistance, is then described.
How angiogenesis as well as the heterogeneous and confined environment
in which a tumor grows is incorporated in the CA model is discussed.
The characterization of  the level of organization of the invasive network
around a solid tumor using spanning trees is subsequently described.
Then, we describe open problems and future promising avenues for future research,
including the need to develop better molecular-based models that
incorporate the true heterogeneous environment over wide range
of length and time scales (via imaging data), cell motility, oncogenes,
tumor suppressor genes and cell-cell communication.
A discussion about the need to bring to bear the powerful machinery
of the theory of heterogeneous media  to better understand the
behavior of cancer in its microenvironment is presented.
Finally, we propose the possibility of using
optimization techniques, which have been used
profitably to understand physical phenomena, in order 
to devise therapeutic (chemotherapy/radiation )
strategies and to understand tumorigenesis itself.
\end{abstract}

\pacs{87.17.Aa, 87.19.xj}
\vspace{2pc}
\noindent{\it Keywords}: tumor growth, glioblastoma multiforme, cellular automaton, heterogeneous media, optimization
\maketitle

\section{Introduction}


Cancer is not a single disease, but rather a highly complex and heterogeneous set of diseases.
Dynamic changes in the genome, epigenome, transcriptome and proteome that result in the gain
of function of oncoproteins or the loss of function of tumor suppressor
proteins underlie the development of all cancers.  While some of  the mechanisms that
govern the transformation of normal cells into malignant ones are rather well
understood \cite{hanahan00}, 
many mechanisms are either not fully understood or are unknown at the moment.
Even if all of the mechanisms could be identified and comprehended,
it is not clear progress in understanding cancer could be made
without knowledge of how these different mechanisms couple
to one another. It has been observed that many complex interactions occur between
tumor cells, and between a cancer and the host environment.  Multidirectional
feedback loops occur between tumor cells and the stroma, immune cells,
extracellular matrix and vasculature 
\cite{ fred91, deisboeck01,brand00,kitano04},
which are not well understood synergistically.
Clearly, our current state of knowledge
is insufficient to deduce clinical outcome, not to mention
how to control cancer progression in the most malignant
forms of cancer.

This suggests that a more quantitative approach to understanding
different cancers is necessary in order to control it and increase
the lifetime of patients with these deadly diseases.
Theoretical/computational modeling of cancer when appropriately linked
with experiments and data offers a promising avenue for
such an understanding. Such modeling of tumor growth using
a variety of different approaches has been a very active
area of research for the last two decades or so  \cite{Ch93,Tr95,Ga96,And98,Pa98,kansal00a,kansal00b,schmitz02,mcdougall02, alarcon05,cristini05,frieboes06,gevertz06,Sw07,gevertz08,gevertz09,Anderson09a, Anderson09b, Fr10} but clearly
is in its infancy. A diverse number of mechanisms have been explored via
such models, and a multitude of computational/mathematical
techniques have been employed; see Ref.~\cite{Byrne10} for a review.
These models have the common aim of predicting certain features
of tumor growth in the hope of finding new ways to
control neoplastic progression. Given a model
which yields reproducible and accurate predictions, the effects
of different genetic, epigenetic and environmental changes,
as well as the impact of therapeutically targeting different
aspects of the tumor, can be probed.
However, these models must  be linked to data from experimental
assays in a comprehensive and systematic fashion in order to
develop of a quantitative understanding of cancer.

The holy grail of computational tumor
modeling is to develop a simulation tool that can be utilized in the clinic to
predict neoplastic progression and response to treatment.  Not only must such a
model incorporate the many feedback loops involved in neoplastic progression,
the model must also account for the fact that cancer progression involves
events occurring over a range of spatial and temporal scales \cite{alarcon05}. 
A successful model would enable one to broaden the conclusions drawn from
existing medical data, suggest new experiments, test new hypotheses, predict
behavior in experimentally unobservable situations, and be employed for early
detection.

There is overwhelming evidence that cancer of all types are
\textit{emerging, opportunistic systems} \cite{coffey98}. Success in treating various cancers
as a self-organizing complex dynamical systems will require unconventional, innovative
approaches and the combined effort of an interdisciplinary team of researchers.
A lofty long-term goal of such an endeavor is
not only to obtain a quantitative understanding of  tumorigenesis
but to {\it limit} and \textit{control}
the expansion of a solid tumor mass and the infiltration of cells
from such masses into healthy tissue.

\centerline{Figure 1: Picture of an Ising model.}

Because a comprehensive
review of the  vast literature concerning biophysical cancer
modeling  is beyond the scope of this
article, we focus  on reviewing the work that we have done toward
the development of an ``Ising model'' of cancer. The Ising model
is an idealized statistical-mechanical model of ferromagnetism
that is based on simple local-interaction rules (see Figure
1), but nonetheless leads to basic insights and features
of real magnets, such as phase transitions with a critical point.
Toward the goal of developing an analogous Ising model of cancer,
we have formulated a four-dimensional (4D) (three dimensions
in space and one in time) cellular automaton (CA) model
for brain tumor growth dynamics and its treatment
\cite{kansal00a,kansal00b,schmitz02,gevertz06,gevertz08,gevertz09}.
Like the Ising model for magnets, we will see later that this
involves local rules for how healthy cells transition into various
types of cancer cells. Before describing our computational models
for tumor growth, we first briefly summarize several salient
features of solid tumor growth as applied to glioblastoma
multiforme (GBM), the most malignant of brain cancers.

The rest of paper is organized as follows: In Section 2,
some background concerning GBMs and solid tumors
in general is presented. In Section 3,
a minimalist 4D CA tumor growth model is described
 in which healthy cells transition between states (proliferative, hypoxic, and necrotic)
according to simple local rules and their present states, and apply it 
to GBMs.
This is followed by a discussion of the extension of the model
to study the effect on the tumor
dynamics and geometry of a mutated subpopulation.
How tumor growth is affected by chemotherapeutic
treatment is also discussed, including induced resistance. 
In Section 4,
the modification of the CA model to include explicitly the effects of vasculature evolution and angiogenesis on tumor growth are discussed.
In Section 5, the effects of physical confinement
and heterogeneous environment are described. In Section 6,
a simulation tool for tumor growth that merges and improves individual CA
models is presented. In Section 7,
a descriptions if given of how one might characterize the invasive network organization
around a solid tumor using spanning trees. Section 8
discusses
some open problems and promising avenues for future research.
 

\section{GBM and Solid Tumor Background}

\centerline{Figure 2: The picture of a tumor in brain.}

Glioblastoma multiforme (GBM) (see Figure 2),   the most
aggressive of the gliomas,  is a collection of tumors arising from the
glial
cells or their precursors in the central nervous system \cite{holland00}.
Unfortunately, despite advances made in cancer biology, the median survival time for
a patient diagnosed with GBM is only 12-15 months, a fact 
that has not changed significantly over the past several decades \cite{Wh10}. 
As suggested by its name (i.e., ``multiforme''), GBM is complex at many
levels of organization \cite{holland00}. 
It exhibits diversity at the macroscopic level, having necrotic, hypoxic
and proliferative regions.  At the mesoscopic level, tumor cell
interactions, microvascular remodeling \cite{holash99a}
and pseudopalisading necrosis are observed \cite{brat04}.
Further, the discovery that tumor stem cells may be the sole malignant
cell type with the ability to proliferate, self-renew and differentiate
introduces yet another level of mesoscopic complexity to GBM \cite{singh04a, singh04b}.
At the microscopic level, GBM cells
exhibit point mutations, chromosomal deletions, and chromosomal
amplifications \cite{holland00}.

\centerline{Figure 3: The picture of a MTS.}

A substantial amount of research has been conducted to model macroscopic tumor growth
either based on microscopic considerations \cite{Du85, Qi93, Sm93};
or in a more phenomenological fashion \cite{Ch93,Wa96}.
One of the early attempts 
to  model empircally the volume $V$ of
a solid tumor versus time $t$ is the Gompertz model, i.e.,
\begin{equation}
V = V_0\exp\left(\frac{A}{B}\left[1-\exp(-Bt) \right] \right),
\end{equation}
where $V_0$ is the volume at time $t=0$ and $A$ and $B$ are
parameters; see Ref.  \cite{steel97} and references
therein. Qualitatively, this equation gives
exponential growth at small times which then saturates at large
times (decelerating growth). In particular, this model considers
the tumor as an oversized idealized multicellular tumor spheroid
(see Figure 3), which is stage of early tumor growth. We
note that modeling an ideal tumor as an oversized spheroid is
especially suited for GBM, since this tumor, like a large MTS,
comprises large areas of central necrosis surrounded by a rapidly
expanding shell of viable cells (Figure 2). However, we
note that real tumors always possess much more complex morphology.
More importantly, Gompertzian-growth models are very limited;
they only capture gross features of tumor growth and cannot
explain their underlying ``microscopic" mechanisms.

One of the hallmarks of high-grade malignant neuroepithelial tumors,
such as glioblastoma multiforme (GBM), is the regional heterogeneity,
i.e., the relatively large number of clonal strains or subpopulations
present within an individual tumor of monoclonal origin \cite{Be92, Co93, Pa89}.
Each of these strains is characterized by
specific properties, such as the rate of division or the level of
susceptibility to treatment \cite{Yu82}.  Therefore the growth
dynamics of a single tumor are determined by the relative behaviors of
each separate subpopulation.  For example, the appearance of a
rapidly dividing strain can substantially bias tumor growth in that
direction.  Tumors supposedly harbor cells with an increased mutation
rate, which indicates that these tumors are genetically unstable
\cite{No76, Lo94, Le98}.
Genetic and epigenetic events throughout the tumor may occur
randomly or be triggered by environmental and intrinsic stresses.  The
continuing existence of a subpopulation, however, depends primarily on
the subpopulation's ability to compete with the dominant population
in its immediate vicinity.

Clonal heterogeneity within a tumor has been shown to have very
pronounced effects on treatment efficacy \cite{He89, Sc86}.
Treatment resistance is itself a complex phenomenon.
There is no single cause of resistance, and many biochemical aspects
of resistance are poorly understood.  Chemoresistant strains can either be resistant
to a single drug or drug family (individual resistance), or they can be resistant
to an array of agents (multidrug/pleotropic resistance) \cite{Br01}.  Cellular mechanisms
behind multidrug resistance include increased chemical efflux and/or decreased
chemical influx, such as with P-glycoprotein-mediated (P-gp) drug
resistance \cite{En89, Ge96}.

Complicating the situation further, resistance can arise at variable times
during tumor development.  Some tumors are resistant to chemotherapy from the onset.
This has been described as {\it inherent} resistance, because it exists before
chemotherapeutic drugs are ever introduced.  In other cases, however, treatment
initially proves successful, and only later does the tumor prove resistant.  This
is an example of acquired resistance, as it develops in
response to treatment \cite{Br01}.  There are at least two possible mechanisms for this
type of tumor behavior.  Acquired resistance may result from a small number of
resistant cells that are gradually selected for throughout the course of treatment.
At the same time, there is also evidence suggesting that chemotherapeutic agents
may induce genetic or epigenetic changes within tumor cells, leading to a resistant
phenotype \cite{Po97}.
Other studies indicate that chemotherapy may increase cellular levels of P-gp mRNA and
protein in various forms of human cancer \cite{Ch93b, Ge94}.
A tumor's response to radiation therapy
can also depend on  underlying genetic factors.  A cell's inherent
radio-resistance may stem from the efficiency of DNA repair mechanisms in sublethally
damaged cells \cite{Ge77, Ka91, Zh93}.

While the properties of GBM cells are very important in understanding growth
dynamics, just as important are the properties of the environment in which the
tumor grows. For example, GBMs grow in either the brain or spinal cord, and are
therefore confined by the shape and size of these organs \cite{gevertz08}. 
Another example of the
importance of accounting for the host environment relates to the vascular
structure of the brain.

Recent research evidence suggests that tumors arising in vascularized
tissue such as the brain do not originate avascularly \cite{holash99a},
as originally thought.  Instead, it is hypothesized that
glioma growth is a process involving vessel co-option, regression and
growth.  Three key proteins, vascular endothelial growth factor (VEGF) and
the angiopoietins, angiopoietin-1 (Ang-1) and angiopoietin-2 (Ang-2), are required to
mediate these processes \cite{holash99a}.
 A picture of what likely occurs during the process of
glioma vascularization has been summarized by Gevertz and Torquato \cite{gevertz06}.
As a malignant mass grows, the tumor cells
co-opt the mature vessels of the surrounding brain that express
constant levels of bound Ang-1.  Vessel co-option leads to the
upregulation in Ang-2 and this shifts the ratio of bound Ang-2 to
bound Ang-1.  In the absence of VEGF, this shift destabilizes the
co-opted vessels within the tumor center and marks them for regression
\cite{holash99b, maisonpierre97}.
Vessel regression in the absence of vessel growth leads to the
formation of hypoxic regions in the tumor mass.  Hypoxia induces the
expression of VEGF, stimulating the growth of new blood vessels \cite{secomb00}.
This robust angiogenic response eventually
rescues the suffocating tumor. Glioma growth dynamics remain
intricately tied to the continuing processes of vessel regression and growth. 


Tumor cell invasion is a hallmark of gliomas \cite{giese96}.
Individual glioma cells have been observed to spread
diffusely over long distances and can migrate into regions of the
brain essential for the survival of the patient \cite{holland00}.
While MRI scans can recognize mass tumor lesions, these scans are not
sensitive enough to identify malignant cells that have spread well
beyond the tumor region \cite{visted03}.  Typically,
when a solid tumor is removed, these invasive cells are left behind
and tumor recurrence is almost inevitable \cite{holland00}. 

Numerous models have been developed to model certain tumor
behavior or characteristics with a great deal of mathematical rigor
(e.g., in the form of coupled differential equations). However, with such approaches, the
sets of equations that govern tumor behavior often do not correspond to the
characteristics of individual tumor cells. An important goal of studying
tumor development is to illustrate how their macroscopic traits stem from their
microscopic properties. In addition, most of the
equations are problem-specific, which limits their utility as general
tools for tumor study. Another potential challenge is that tumor models
should be appreciated by as diverse an audience as possible.  Ideally, the mathematical
complexity that allows theoreticians to analyze subtle aspects of it should
not be an obstacle for clinicians who treat GBM.  A model that accounts
for complex tumor behavior with relative mathematical ease could be valuable.

To this end, we have developed what appears to be a powerful cellular automaton (CA) computational tool
for tumor modeling. Based on a few salient set of microscopic parameters, this CA model can realistically
model the macroscopic tumor behavior, including growth dynamics, emergence of a subpopulation
as well as the effects of tumor treatment and resistance \cite{kansal00a, kansal00b, schmitz02}.
This model has been extended to study the
effects of vasculature evolution on early tumor growth and to simulate tumor growth
in confined heterogeneous environments \cite{gevertz06, gevertz08, gevertz09}.
We have also developed mathematical models to characterize the invasive network organization
around a solid tumor \cite{kansal01}.


\section{Toward an Ising Model of Cancer Growth}

In this section, we describe a four-dimensional (4D) cellular
automaton (CA) model that we have developed that describes tumor
growth as a function of time, using the fewest number of microscopic parameters
\cite{kansal00a, kansal00b, schmitz02}. We refer to this
as a {\it minimalist} four-dimensional (4D) model because it involves three
spatial dimensions and one dimension in time with the goal
of capturing the salient features of tumor growth 
with a minimal number of parameters. The algorithm takes into
account that this growth starts out from a few cells, passes
through a multicellular tumor spheroid (MTS) stage (Figure
3) and proceeds to the macroscopic stages at clinically
designated time-points for a virtual patient: detectable lesion,
diagnosis and death. This 4D CA algorithm models
macroscopic GBM tumors as an enormous idealized MTS,
mathematically described by a Gompertz-function given by Eq. (1),
since this tumor, like a large MTS, comprises large areas of
central necrosis surrounded by a rapidly expanding shell of viable
cells (Figure 2). In accordance with experimental data,
the algorithm also implicitly takes into account that invasive
cells are continually shed from the tumor surface and implicitly assumes that
the tumor mass is well-vascularized during the entire process of
growth. The effects of vasculature evolution are considered
explicitly in  Sections 5 and 7.

\subsection{A 4D Cellular Automaton Model}

A CA model is a spatially and temporally discrete model
that consists of a grid of cells, with each cell being in one of a number of
predefined states.  The state of a cell at a given point in time depends on
the state of itself and its neighbors at the previous discrete time point.
Transitions between states are determined by a set of local rules.
The simulation is designed to predict clinically important
criteria such as the fraction of the tumor which is able to divide
(GF), the non-proliferative ($G_0/G_1$ arrest) and necrotic
fractions, as well as the rate of growth (volumetric doubling
time) at given radii. Furthermore, this CA model enables one to
study emergence of a subpopulation due to cell mutations as
well as the effects of tumor treatment and resistance. The general
CA model includes both a proliferation routine which models tumor
growth by cell division and a treatment routine which models the cell
response to treatment and cell mutations. It also incorporates a novel
adaptive automaton cell generation procedure. In particular, the CA model is characterized by
several biologically important features:

\begin{itemize}
\item The model is able to grow the tumor from a very small size
of roughly 1000 real cells through to a fully developed tumor with
$10^{11}$ cells. This allows a tumor to be grown from almost any
starting point, through to maturity.
\item The thickness of different tumor layers, i.e. the proliferative rim and the
non-proliferative shell, are linked to the overall tumor radius by
a 2/3 power relation.  This reflects a surface area to volume
ratio, which can be biologically interpreted as nutrients
diffusing through a surface.
\item The discrete nature of the model and the
variable density lattice allow us to control the inclusion of
mutant ``hot spots'' in the tumor as well as variable cell
sensitivity/resistance to treatment. The variable
density lattice will allow us to look at such an area at a higher resolution.
\item Our inclusion of mechanical confinement pressure enables us to simulate the physiological
confinement by the skull at different locations within the brain
differently.
\end{itemize}

Our CA algorithm can be broken into three parts: automaton
cell generation, the proliferation routine and the treatment routine.
In the ensuing discussions, we first present the three parts of our algorithm.
Then we show that the our model reflects a test case derived from the medical
literature very well, proving the hypothesis that macroscopic
tumor growth behavior may be modeled with primarily microscopic
data.

\subsubsection{Cellular Automaton Cell Generation}

The first step of the simulation is to generate the automaton
cells. The underlying lattice for the algorithm is the Delaunay
triangulation, which is the dual lattice of the Voronoi
tessellation \cite{kansal00a, torquato02}.  In
order to develop the automaton cells, a prescribed number of
random points are generated in the unit square using the process
of random sequential addition (RSA) of hard circular disks. In the RSA procedure, as
a random point is generated, it is checked if the point falls
within some prescribed distance from any other point already
placed in the system \cite{kansal00a, torquato02}.
Points that fall too close to any other point are
rejected, and all others are added to the system. Each cell in the
final Voronoi lattice will contain exactly one of these accepted
sites.  The Voronoi cell is defined by the region of space nearer
to a particular site than any other site.  In two-dimensions, this
results in a collection of polygons that fill the plane.

\centerline{Figure 4: Voronoi Cells}

Because a real brain tumor grows over several orders of magnitude
in volume, the lattice was designed to allowed continuous
variation with the radius of the tumor. The density of lattice
sites near the center was significantly higher than that at the
edge. A higher site density corresponds to less real cells per
automaton cell, and so to a higher resolution. The higher density
at the center enables us to simulate the flat small-time behavior
of the Gompertz curve. In the current model, the innermost
automaton cells represent roughly 100 real cells, while the
outermost automaton cells represent roughly $10^6$ real cells. The
average distance between lattice sites was described by the
following relation:
\begin{equation}
\zeta=\frac{1}{6}r^{2/3}
\end{equation}
in which $\zeta$ is the average distance between lattice sites and
$r$ is the radial position at which the density is being measured.
This relation restricts the increase in the number of
proliferating cells as the tumor grows. Note that when modeling
the effects of vasculature evolution discussed in the following, a
a uniform lattice is used for which each automaton cell includes
approximately 10 real cancer cells.

\subsubsection{Minimalist 4D Proliferation Algorithm}

\centerline{Figure 5: Idealized tumor}

\begin{table}[h]
\begin{center}
\caption{The time-dependent functions and growth, treatment parameters for the model.}
\begin{tabular}{ll} \hline\hline
\multicolumn{2}{c}{\textbf{Functions within the model (time dependent)}} \\ 
$R_t$ & Average overall tumor radius (see Appendix) \\ 
$\delta_p$ & Proliferative rim thickness (determines growth fraction) \\ 
$\delta_n$ & Non-proliferative thickness (determines necrotic fraction) \\ 
$p_{d}$ & Probability of division (varies with time and position) \\ 
\\
\multicolumn{2}{c}{\textbf{Growth parameters}} \\ 
$p_0$ & Base probability of division, linked to cell-doubling time (0.192) \\ 
$a$ & Base necrotic thickness, controlled by nutritional needs ($0.42\;\mbox{mm}^{1/3}$) \\ 
$b$ & Base proliferative thickness, controlled by nutritional needs ($0.11\;\mbox{mm}^{1/3}$) \\ 
$R_{max}$ & Maximum tumor extent, controlled by pressure response
($38\;\mbox{mm}$) \\ \hline\hline
\end{tabular}
\label{ParamTab}
\end{center}
\end{table}

The proliferation algorithm is designed to allow a tumor
consisting of a few automaton cells, representing roughly 1000
real cells, to grow to a full macroscopic size. An idealized model
of a macroscopic tumor is an almost spherical body consisting of
concentric shells of necrotic, non-proliferative and proliferative
regions (see Figure 5). The four microscopic growth
parameters of the algorithm are $p_0$, $a$, $b$, and $R_{max}$
reflecting, respectively, the rate at which the proliferative
cells divide, the nutritional needs of the non-proliferative and
proliferative cells, and the response of the tumor to mechanical
pressure within the skull. In addition, there are four key
time-dependent quantities that determine the dynamics of the
tumor, i.e., $R_{t}$, $\delta_{p}$, $\delta_n$, $p_d$ giving,
respectively, the average overall tumor radius, proliferative rim
thickness, non-proliferative thickness and probability of
division. These quantities are based on the four parameters
($p_0$, $a$, $b$, $R_{max}$) and are calculated according to the
following algorithm.

\begin{itemize}
\item Initial setup: The cells within a fixed initial radius of
the center of the grid are designated proliferative. All other
cells are designated as non-tumorous.
\item Time is discretized and incremented, so that at each time
step:
\begin{itemize}
\item Each cell is checked for type: non-tumorous or (apoptotic
and) necrotic, non-proliferative or proliferative tumorous cells.
\item Non-tumorous cells and tumorous necrotic cells are inert.
\item Non-proliferative (growth-arrested) cells more than a
certain distance, $\delta_n$, from the tumor's edge are turned
necrotic.  This is designed to model the effects of a nutritional
gradient. The edge of the tumor is taken to be the nearest
non-tumorous cell, i.e.,
\begin{equation}
\delta_{n} = aR_{t}^{2/3}.
\end{equation}
\item Proliferative cells are checked to see if they will attempt
to divide according to the probability of division, $p_{d}$, which
is influenced by the location of the dividing cell, reflecting the
effects of mechanical confinement pressure. This effect requires
the use of an additional parameter, the maximum tumor extent,
$R_{max}$. $p_d$ is given by
\begin{equation}
p_{d}=p_{0}(1-\frac{r}{R_{max}}) \label{presseq}
\end{equation}
\item If a cell attempts to divide, it will search for sufficient
space for the new cell beginning with its nearest neighbors and
expanding outwards until either an empty (non-tumorous) space is
found or nothing is found within the proliferation radius,
$\delta_{p}$.  The radius searched is calculated as:
\begin{equation}
\delta_{p}=bR_{t}^{2/3}.
\end{equation}
\item If a cell attempts to divide but cannot find space it is
turned into a non-proliferative cell.
\end{itemize}
\item After a predetermined amount of time has been stepped
through, the volume and radius of the tumor are plotted as a
function of time.
\item The type of cell contained in each grid are saved at given times.
\end{itemize}

\centerline{Figure 6: An illustration of the
proliferation algorithm.}

The above simulation procedure is also illustrated in Figure
6.  We note that
the redefinition of the proliferative to non-proliferative
transition implemented in the algorithm is one of the most
important new features of the model. They allow a larger number of
cells to divide, since cells no longer need to be on the outermost
surface of the tumor to divide. In addition, it ensured that cells
that cannot divide are correctly labeled as such. Table
\ref{ParamTab} summarizes the important time-dependent functions
calculated by the proliferation algorithm and the constant growth
parameters used. The readers are referred to Ref. \cite{kansal00a}
for the detailed description of the algorithm and parameters.


\subsubsection{Extending the 4D CA Model to Study Emergence of a Subpopulation}

Malignant brain tumors such as GBM generally consist of a number
of distinct subclonal populations. Each of these subpopulations,
arising from the constant genetic and epigenetic alteration of
existing cells in the rapidly growing tumor, may be characterized
by its own behaviors and properties. However, since each single
cell mutation only leads to a small number of offspring initially,
very few newly arisen subpopulations survive more than a short
time.  Kansal \textit{et al} \cite{kansal00b} have extended the CA
to quantify ``emergence,'' i.e. the likelihood of
an isolated subpopulation surviving for an extended period of
time. Only mutations affecting the rate of cellular division were
considered in this rendition of the model. In addition, only competition between
clones was taken into account; there were no cooperative effects
included, although such effects can easily be incorporated.

The simulation procedure is as follows:
an initial tumor composed entirely of cells of
the primary clonal population is introduced,
which is allowed to grow using the proliferation algorithm
until it reaches a predetermined average overall radius. Then, a single (or a small number of)
automaton cell is changed from the \textit{primary} strain to a
\textit{secondary} strain with an altered probability of division,
which represents very small fractions of the total
population of proliferative tumor cells
and the tumor is allowed to continue to grow using the proliferation algorithm.
It is important to note that this does not represent a single mutation
event but rather a mutation event that results in a subpopulation
reaching a size dictated by the limits of the lattice resolution 
employed (i.e., a specified number of cells).

The behavior of the secondary strain was characterized in terms
of two properties: the \textit{degree} $\alpha$ and the \textit{relative size} $\beta$
of the initial population of mutated cells, i.e.,
\begin{equation}
\alpha = \frac{p_1}{p_0},
\end{equation}
\noindent which represents the ratio between the base probability
of division of the new clone, $p_1$, and that of the original clone,
$p_0$; and
\begin{equation}
\beta = \frac{\mbox{volume of proliferating cells of
the new clone}}
{\mbox{volume of proliferating cells of the original clone}}.
\end{equation}
Positive, negative and no competitive advantages are
respectively conferred for $\alpha > 1$, $\alpha<1$, and $\alpha =
1$. The initial value $\beta$, i.e., $\beta_0=\beta(t=0)$, is a
parameter of the model reflecting the size of the mutated region
introduced.


\subsubsection{Extending the 4D CA Model to Study Treatment}

Besides the four growth parameters in the minimalist 4D CA model, three additional parameters
for treatment  were subsequently introduced: $\gamma$, $\epsilon$,
and $\phi$, the values of which reflect, respectively, the
proliferative cells' treatment sensitivity, the non-proliferative
cells' treatment sensitivity, and the mutational response of the
tumor cells to treatment \cite{schmitz02}. Furthermore, there are three additional
time-dependent quantities $D_{pro}$, $D_{non}$ and $\beta$, giving
respectively fraction of proliferative cells that die upon
treatment (equivalent to $\gamma$), fraction of non-proliferative
cells that die upon treatment (equivalent to $\gamma\epsilon$) and
volume fraction of mutated living cells. These parameters are
summarized in Table 2 and a detailed discussion is given in
Ref. \cite{schmitz02}.

\begin{table}[h]
\begin{center}
\caption{Treatment parameters and associated terms for the model.}
\begin{tabular}{ll} \hline\hline
\multicolumn{2}{c}{\textbf{Treatment parameters}} \\ 
$\gamma$ & Governs the proliferative cells' response at each
\\ & instance of treatment ($0.55-0.95$) \\ 
$\epsilon$ & Allows for different treatment responses between
\\ & proliferative and non-proliferative cells ($0-0.4$)\\ 
$\phi$ & Fraction of surviving proliferative cells
\\ & that mutate in response to treatment ($10^{-5}-10^{-2}$) \\ 
\\
\multicolumn{2}{c}{\textbf{Other Terms}} \\ 
$D_{pro}$ & Fraction of proliferative cells that die
\\ & upon treatment; equivalent to $\gamma$ \\ 
$D_{non}$ & Fraction of non-proliferative cells that die
\\ & upon treatment; equivalent to $\gamma\epsilon$ \\ 
$\beta$ & Volumetric fraction of living cells (proliferative and
\\ & non-proliferative) belonging to the secondary strain \\ \hline\hline
\end{tabular}
\label{TreatTab}
\end{center}
\end{table}

In the simulation, treatment was introduced as ``periodic
impulse'', i.e., a small tumor mass is introduced which is
intended to represent a GBM after successful surgical resection
and allowed to grow using the proliferation algorithm; then
treatment is applied and considered effective at discrete time
points. In particular, the simulation proceeds through the
proliferative steps until every $n_w$ week time-point, at which
time the treatment routine is introduced:

\begin{itemize}
\item After the last round of cellular division, each
proliferative cell is checked to see if it is killed by the
treatment. The probability of death for a given proliferative cell
$D_{pro}$ is given by
\begin{equation}
D_{pro} = \gamma,
\end{equation}
\noindent where $\gamma \in (0,~1)$ is the proliferative treatment
parameter. Dead proliferative automaton cells are converted to
healthy cells. \item Each non-proliferative cell is checked to see
if it is killed.  The probability of death for a given
non-proliferative cell $D_{non}$ is given by
\begin{equation}
D_{non} = \gamma\epsilon,
\end{equation}
\noindent where $\epsilon\in(0,~1)$ is the non-proliferative
treatment parameter and $D_{non}$ is a fraction of $D_{pro}$. A
non-proliferative cell is converted to a necrotic cell upon death.
\item Each surviving non-proliferative cell is checked to see if
it is within the proliferative thickness of a healthy cell (i.e.
the tumor surface). If so, the non-proliferative cell is converted
back to a proliferative cell. \item All proliferative cells
(including newly-designated ones) are checked for mutations for
treatment resistance $\gamma$ with probability $\phi$. A new
$\gamma\in(0,~1)$ is randomly generated for mutated cells while
$\epsilon$ remains constant.
\end{itemize}

Clinically, GBM treatment consists of both
radiation therapy and chemotherapy. However, in our model we do
not distinguish between the separate effects of these two methods.
The tumors' response to all treatment is captured by the treatment
algorithm. Moreover, this response is assumed to be instantaneous
at each four-week time point.


\section{Putting the 4D CA Model Through Its Paces}

\subsection{A Test Case for Proliferation Algorithm}

The tumor growth data generated via the minimalist 4D CA proliferation algorithm
was compared with available experimental data for an
untreated GBM tumor from the medical literature \cite{kansal00a}.
The parameters compared were cell number, growth
fraction, necrotic fraction and volumetric doubling time, which
are used to determine a tumor's level of malignancy and the
prognosis for its future growth. Because it is impossible to
determine the exact time a tumor began growing, the medical data
are listed at fixed radii. The different cell fractions used were
extrapolated from the spheroid level and compared to data
published for cell fractions at macroscopic stages.

\begin{table}[h]
\label{CompTab} \centering \caption{ Comparison of simulated tumor
growth and experimental data. For each quantity, the simulation
data is give on the first line and the experimental data is given
on the second line.}
\begin{tabular}{ll} \hline\hline
$\begin{array}{c@{\hspace{2.5mm}}c@{\hspace{2.5mm}}c@{\hspace{2.5mm}}c@{\hspace{2.5mm}}c}
  & \mbox{Spheroid} & \mbox{Det. Les.} & \mbox{Diagnosis} & \mbox{Death} \\ \hline
\mbox{Time} & \mbox{Day}~69 & \mbox{Day}~223 & \mbox{Day}~454 & \mbox{Day}~560 \\ \\
\mbox{Radius} & 0.5\; \mbox{mm}&  5\; \mbox{mm}&  18.5\; \mbox{mm}& 25\; \mbox{mm}\\
      & 0.5\; \mbox{mm}&  5\; \mbox{mm}&  18.5\; \mbox{mm}& 25\; \mbox{mm}\\ \\
\mbox{Cell No.} & 10^6 & 10^9 & 5\times 10^{10} & 10^{11} \\
    & 7\times 10^5 & 6 \times 10^8 & 4\times 10^{10} & 9\times 10^{10} \\ \\
\mbox{Growth fraction} & 0.36 & 0.30 & 0.20 & 0.09 \\
    & 0.35 & 0.30 & 0.18 & 0.11 \\ \\
\mbox{Necrotic fraction} & 0.46 & 0.49 & 0.55 & 0.60 \\
    & 0.38 & 0.53 & 0.58 & 0.63 \\ \\
\mbox{Volume-doubling time} & 6~\mbox{days} &  45~\mbox{days} &  70~\mbox{days} &  105~\mbox{days} \\
& 9~\mbox{days} &  36~\mbox{days} &  68~\mbox{days} &  100~\mbox{days} \\ \hline\hline
\end{array}$
\end{tabular}
\end{table}

Summarized in Table 3 is the comparison between
simulation results and data (experimental, as well as clinical)
taken from the medical literature (see Ref. \cite{kansal00a} for
detailed references). The simulation data were created using a
tumor which was grown from an initial radius of 0.1 mm.  The
following parameter set (see Table 1) was used:

\centerline{$p_0 = 0.192$, $a = 0.42\; \mbox{mm}^{1/3}$, $b = 0.11\;
\mbox{mm}^{1/3}$, $R_{max} = 37.5\; \mbox{mm}$}

\noindent This value of $p_0$ corresponds to a cell-doubling time
of 4.0 days. The parameters $a$ and $b$  have been chosen to give a
growth history that quantitatively fits the test case. The
specification of these parameters corresponds to the specification
of a clonal strain. The parameter $R_{max}$  was similarly chosen
to match the test case history. In this case, however, the fit is
relatively insensitive to the value of $R_{max}$, as long as the
parameter is somewhat larger than the fatal radius in the test
case. On the whole, the simulation data reproduces the test case
very well. The virtual patient would die roughly 11 months after
the tumor radius reached 5 mm and 3.5 months after the expected
time of diagnosis.  The fatal tumor volume is about $65 \;\mbox{cm}^3$.

\centerline{Figure 7: Cross-sections of a Growing Mono-Clonal
Tumor.}

\centerline{Figure 8: A cut-away view of the simulated
tumor.}

\centerline{Figure 9: The Volume and Radius of the
Developing Tumor.}

Central cross-sections of the tumors are shown in Figure
7, in which the growth of the tumor can be followed
graphically over time. Here necrotic cells are labeled with black,
non-proliferative tumorous cells with light gray and proliferative
tumor cells with dark gray. { A cut-away view of the simulated
tumor is shown in Figure 8.} As expected in this
idealized case, the tumor is essentially spherical, within a small
degree of randomness. The high degree of spherical symmetry
ensures that the central cross-section is a representative view.
The volume and radius of the developing tumor are shown versus
time in Figure 9. Note that the virtual patient dies
while the untreated tumor is in the rapid growth phase.

\subsection{Modeling the Emergence of A Subpopulation}


Recall that the parameter $\alpha$ reflects
the degree of advantage of the mutated subpopulation over 
the primary clone (positive, negative and no competitive advantages are
respectively conferred for $\alpha > 1$, $\alpha<1$, and $\alpha =
1$) and  the initial value $\beta$, i.e., $\beta_0=\beta(t=0)$, is a
parameter of the model reflecting the size of the mutated region
introduced.
A subpopulation is considered to have emerged once it comprises
$5\%$ of the actively dividing cell population or if it remains in
the actively dividing state once the tumor has reached a fully
developed size. Numerous simulations (at least 100) were run at
each parameter set by Kansal \textit{et al} \cite{kansal00b} in order to
calculate the expected probabilities of emergence, i.e.,
\begin{equation}
P = \frac{\mbox{number of trials in which emergence occurs}}
{\mbox{total number of trials}}.
\end{equation}
along with confidence intervals, $\sigma$, defined as
\begin{equation}
\sigma = \sqrt{\frac{\overline{p}(1-\overline{p})}{N}}
\end{equation}
\noindent where $\overline{p}$ represents the observed probability
of emergence in $N$ trials. We note that the probability of
emergence is actually a conditional probability: it is the
probability that a subpopulation with a mutation of degree
$\alpha$ emerges given that a region of relative size $\beta_0$
has mutated.

\centerline{Figure 10: P vs. alpha and cut-away view
of simulated tumor with a subpopulation.}

The results represented were run with a parameter set in which

\centerline{$p_0=0.192$, $p_1=\alpha p_0$, $a=0.42\;\mbox{mm}^{1/3}$,}
\centerline{$b=0.11\;\mbox{mm}^{1/3}$, $R_{max} = 37.5\;\mbox{mm}$}

\noindent for the primary strain, in a simulation in which each
time step represents one day \cite{kansal00b}. Figure 10
depicts the observed probability of emergence, $P$, for a
subpopulation of initial size $\beta_0=6\times10^{-5}$ as a
function of $\alpha$, which gives an approximation of the true,
asymptotic, probability of emergence. Also shown in Figure
10  is a cut-away view of the simulated tumor with a
subpopulation. Not surprisingly, $P$ is a monotonically
increasing function that tends to 0 for $\alpha < 1$ and to 1 as
$\alpha$ become significantly greater than 1. 

Perhaps the most
striking feature of these results is that there is a non-zero
probability of emergence for a very small population with no
growth rate advantage, or even with a small disadvantage (i.e.
$\alpha \approx 0.95$). This suggests that a mutated subpopulation
may arise even without any growth advantage. These populations
could represent ``dormant'' clones which confer an advantage not
being selected for at the time.  An example would be the
appearance of hypoxia tolerant or even treatment resistant clones.
It should be stressed that  populations with less competitive
advantages over other tumor strains can have a
nonzero probability of emergence especially if they are {\it localized} in space, which
leads to a minimum surface area between the two populations per
unit tumor volume. In this way, the  population with smaller
competitive advantage can compete more effectively. We will see
in the next subsection that this same principle is 
at work when resistance is induced due to treatment. 
It was also found that the emergence probability $P$
is a monotonically increasing function in $\beta_0$ and has a
logarithmic dependence on $\beta_0$ \cite{kansal00b}.

\centerline{Figure 11: Effects of the subpopulation on the
tumor geometry.}

\centerline{Figure 12: Effects of the subpopulation on
growth history.}

Figure 11 shows the effects of growth of the
subpopulation on the tumor geometry. It can be seen clearly that
the center of mass of the tumor is significantly shifted by the
emergence of the subpopulation. Another example of the importance
of subpopulations is depicted in Figure 12
\cite{kansal00b}. In this example, a diagnosis was made (on
day $t_0$) giving information about the macroscopic size and
growth rate of the tumor. From this information three possible
growth histories of the tumor are plotted.  One is the time
history of the tumor with an emergent subpopulation. The others
represent limiting cases, each with a monoclonal tumor of either
the primary (``base $p_0$'') or secondary (``high $p_0$'') clonal
strain. Note that at the time of diagnosis all three scenarios
have very similar dynamics.  So any of the three histories is a
reasonable prediction given only size and growth rate information.
However, estimating a fatal tumor volume to be 65 $\mbox{cm}^3$ and
defining the survival time to be the time required to reach this
volume, the base case mis-predicts survival times to be 90 days, which
is 30 days more than the 60 days of the ``true'' course. 

It is
noteworthy that from this perspective the overall future growth
dynamics of the entire tumor closely follows that of the most
aggressive case, indicating that the more aggressive clone
dominates overall outcome and should therefore also define the appropriate
treatment. This finding supports the current practice in pathology
of grading tumors according to the most malignant area (i.e.
population) found in any biopsy material. Although of less
clinical significance, the high case similarly mis-predicts the
past history of the tumor. If the diagnosis had been made earlier,
the base case would yield still worse future predictions.
Similarly the ``high" $p_0$ case would yield worse past predictions for a
diagnosis made at a later time. The predictive errors arising from
the assumption of a monoclonal tumor indicate how important an
accurate estimate of the clonal composition of a tumor is in
establishing a complete history and prognosis. Note that the
numbers given here are intended to show the scale of the
inaccuracy possible, not to reflect any data extracted from actual
patients.


\subsection{Modeling the Effects of Tumor Treatment and Resistance}

Combining the proliferation algorithm and the treatment algorithm,
the behavior of tumors that are able to develop resistance
throughout the course of treatment were investigated \cite{schmitz02}.
Recall that additional parameters were
introduced in the treatment routine: $\gamma$, $\epsilon$, and
$\phi$, the values of which reflect, respectively, the
proliferative cells' treatment sensitivity, the non-proliferative
cells' treatment sensitivity, and the mutational response of the
tumor cells to treatment (see Table 2). 

These investigation consisted of three individual case studies.
In Case 1, the growth dynamics of monoclonal
tumors are studied to determine how tumor behavior is affected by the treatment
parameters $\gamma$ and $\epsilon$.  Case 2 builds upon this information, analyzing
the behavior of two-strain tumors.  Here, a secondary treatment-resistant strain
exists alongside a primary treatment-sensitive strain.  A
secondary sub-population was introduced at the onset of each simulation, 
initializing it in different
spatial arrangements and at several (small) relative volumes.  In both Cases 1 and 2, no
additional sub-populations arise in the tumors once the simulation has begun (i.e. $\phi = 0$).
In Case 3, however, tumors were studied that were capable of undergoing
resistance mutations in response to each round of treatment ($\phi > 0$).  In these
simulations, the growth and morphology of the tumors were analyzed in relation to the
fraction of mutating cells. 

Here we only report on the results of Cases 2 and 3.
In Case 2, the smaller subpopulation of a secondary treatment-resistant strain
was initially spatially distributed in two different ways on the tumor
surface that primarily consist of the primary treatment-sensitive strain:
a {\it localized} and {\it scattered} scenario, reflecting
possible effects of the result of surgery, for example (see Figure 13).
In the simulation, the
tumors were initialized as a single strain, i.e., monoclonally with
$\gamma = 0.95$ and $\epsilon = 0.05$ and treatment was introduced
every four weeks while the tumor is growing from a small mass with
a radius of 4mm, corresponding to approximately $99\%$ of surgical
volume resection.
For the scattered resistance scenario, the resistant strain was 
found to compete more effectively with the sensitive strain and
it was shown that the initial number of resistant cells 
were not a significant indicator or prognosis. 

\centerline{Figure 13: Spatial distributions of the resistance
strain.}

These conclusions may at first glance seem to contradict the those
reported by by Kansal \textit{et al} \cite{kansal00b}. Recall
that in this work the selection pressure was different (growth-rate competition versus treatment effects). Moreover, the roles of the primary and secondary strains are reversed in the
Case 2 example: the primary strain possessed a
competitive advantage over the secondary strain.  Nevertheless, the conclusions
of both papers \cite{kansal00b,schmitz02} follow precisely the
same principle. The proliferative ability of a strain with a competitive advantage varies
directly with its contact area with the less comptetive strain per cell.

Unlike Case 2, the tumors in Case 3 begin
the simulations as a single strain.  Here, however, treatment can induce the
appearance of mutant strains ($\phi > 0$).  In these
simulations, the growth and morphology of the tumors were analyzed in relation to the
fraction of mutating cells. The tumors in Case 3 are all initialized monoclonally with 
$\gamma = 0.95$ and $\epsilon = 0.05$. With this initial $\gamma$-value, nearly 
every mutant strain that arises from the initial population will posses a 
lower $\gamma$-value. This is not to suggest that all induced
mutations must possess increased resistance. This fact here merely stems from 
the initial sensitive tumor under consideration.

At first, the tumors in Case 3 will develop like treatment-sensitive, 
monoclonal tumors; growth will then accelerate as resistant cells begin 
to dominate.  This corresponds to a case of acquired resistance via induced 
(genetic and epigenetic) mutations.  Overall, the tumor dynamics here are more
variable than in Cases 1 and 2.  When a new strain appears, it begins as a 
single automaton cell.  Unlike Case 2, not all new strains will be able to 
proliferate to an appreciable extent.  Some are
overwhelmed by the parent strain from which they arise.

The mean survival time of the tumors were determined as a function of $\phi$
and these data are summarized in  Figure 14. plots this data;
From $\phi=10^{-5}$ to $\phi=10^{-2}$, the survival times vary nearly logarithmically with $\phi$.
When $\phi=10^{-5}$, the mean time is near 27 months, as most tumors remain
monoclonal (or nearly monoclonal) with $\gamma = 0.95$, $\epsilon = 0.05$.  As $\phi$
increases, resistant strains appear more commonly and survival times fall.

\centerline{Figure 14: Survival times associated with continuously mutating tumors.}

One of the more intriguing observations in this case involves
the gross morphology of the mutating tumors. Their
three-dimensional geometries exhibit an interesting dependence on
the value of $\phi$. Figure 15 presents representative
images of the fully-developed tumors for small, intermediate and
large fractions of mutated proliferative cells $\phi$ after
treatment. For small $\phi$ (left panel of Figure 15), some
tumors develop a secondary strain while others do not. The tumors
that remain monoclonal maintain their spherical geometry.  When a
resistant sub-population does develop, it appears as a lobe on the
parental tumor.  For intermediate $\phi$, resistant
sub-populations consistently arise from the parental strain.  The
middle panel of Figure 15 depicts a typical tumor,
whose geometries consistently deviate from an ideal sphere. These
tumors are multi-lobed in appearance, and the original strain is
commonly overwhelmed. However, when $\phi$ is large, the geometric
trend reverses, i.e., the tumors (right panel of Figure 15) 
again appear more spherical, despite the fact that they
experience the greatest fraction of mutations per treatment event.
These images suggest that extreme mutational responses can lead to
similar macroscopic geometries.  Non-spherical geometries result
from intermediate $\phi$-values.

\centerline{Figure 15: Images of  continuously mutating tumors.}

\section{Modeling the Effects of Vasculature Evolution}

As pointed out in the Introduction, there are complex interactions
occurring between between a tumor and the host environment, which
makes it very difficult in predicting clinical outcome, even
if mutations responsible for oncogenesis that determine tumor growth
are beginning to be understood. These interactions include
the effects of vasculature evolution on tumor growth, the
organ-imposed physical confinement as well as the host
heterogeneity. While the three studies described in the previous
section were successful at analyzing and characterizing 
GBM growth both with and without treatment, in each case, the
CA model made the simplifying assumption that
the tumor mass was well-vascularized (the vascular network and angiogenesis
were implicitly accounted for) and the effects of mechanical
confinement were limited to one parameter ($R_{max}$), which allowed
for growth of  spherically symmetric tumors with a maximum radius.
Spherical-like growth is realistic provided that the environment
is effectively homogeneous, but heterogeneous environments
will cause apsherically-shaped tumors.

In order to incorporate a greater level of  microscopic detail, a
3D  (two dimensions in space and one in time) hybrid variant of the original CA model that
allows one to study how changes in the tumor vasculature due to
vessel co-option, regression and sprouting influence GBM
was developed by Gevertz and Torquato \cite{gevertz06}.
This computational algorithm is based
on the co-option-regression-growth experimental model of tumor
vasculature evolution \cite{holash99a, holash99b}.  In this
model, as a malignant mass grows, the tumor cells co-opt the
mature vessels of the surrounding tissue that express constant
levels of bound angiopoietin-1 (Ang-1). Vessel co-option leads to
the upregulation of the antagonist of Ang-1, angiopoietin-2
(Ang-2). In the absence of the anti-apoptotic signal triggered by
vascular endothelial growth factor (VEGF), this shift destabilizes
the co-opted vessels within the tumor center and marks them for
regression \cite{holash99a, holash99b}. Vessel regression in
the absence of vessel growth leads to the formation of hypoxic
regions in the tumor mass. Hypoxia induces the expression of VEGF,
stimulating the growth of new blood vessels.

A system of reaction-diffusion equations was developed to track
the spatial and temporal evolution of the aforementioned key
factors involved in blood vessel growth and
regression \cite{gevertz06} (see Section 6 for a detailed description).
Based on a set of algorithmic rules, the
concentration of each protein and bound receptor at a blood vessel
determines if a vessel will divide, regress or remain stagnant.
The structure of the blood vessel network, in turn, is used to
estimate the oxygen concentration at each cell site.  Oxygen
levels determine the proliferative capacity of each automaton
cell.  The reader is referred to \cite{gevertz06} for
the full details of this algorithm.  The model proved to
quantitatively agree with experimental observations on the growth
of tumors when angiogenesis is successfully initiated and when
angiogenesis is inhibited.  Further, due to the biological details
incorporated into the model, the algorithm was used to explore
tumor response to a variety of single and multimodal treatment
strategies \cite{gevertz06}.

\section{Modeling the Effects of Physical Confinement and Heterogeneous Environment}

An assumption made in both the original CA algorithm and the one that
explicitly incorporates 
vascular evolution is that the tumor is growing in a spherically
symmetric fashion. In a study performed by Helmlinger \textit{et al} \cite{helmlinger97},
it was shown that neoplastic growth
is spherically symmetric only when the environment in which the
tumor is developing imposes no physical boundaries on growth. In
particular, it was demonstrated that human adenocarcinoma cells grown in
a 0.7\% gel that is placed in a cylindrical glass tube develop to
take on an ellipsoidal shape, driven by the geometry of the
capillary tube.  However, when the same cells are grown in the
same gel outside the capillary tube, a spherical mass develops
\cite{helmlinger97}. This experiment clearly
highlights that the assumption of radially symmetric growth is
only valid when a tumor grows in an unconfined or spherically
symmetric environment.

Since many organs, including the brain and spinal cord, impose
non-radially symmetric physical confinement on tumor growth,
the original CA algorithm was modified to incorporate boundary and
heterogeneity effects on neoplastic progression \cite{gevertz08}.
The first modification that was made
to the original algorithm was simply to specify and account for the boundary that is
confining tumor growth. Several modifications
were made to the original automaton rules to account for the
impact of this boundary on neoplastic progression.  The original
CA algorithm imposed radial symmetry in order to determine whether
a cancer cell is proliferative, hypoxic, or necrotic. The
assumption of radially symmetric growth was also utilized in
determining the probability a proliferative cell divides.  In
order to allow tumor growth in any confining environment,
all assumptions of radial symmetry from the automaton
evolution rules were removed. It was
 demonstrated that models that do not account for the geometry of
the confining boundary and the heterogeneity in tissue structure
lead to inaccurate predictions on tumor size, shape and spread
(the distribution of cells throughout the growth-permitting
region). The readers are referred to Ref. \cite{gevertz08} for 
the details of this investigation, but an illustration
of confinement effects are given in the next section.

\section{A Merged Tool for Growing Heterogeneous Tumors In Silico}

\subsection{Algorithmic Details}

Each of the previously discussed algorithms were designed to
answer a particular set of questions and successfully served their
purpose. Hence,  Gevertz and Torquato \cite{gevertz09} merged 
each algorithm into a single cancer simulation tool that would not only accomplish what
each individual algorithm had accomplished, but had the capacity
to have emergent properties not identifiable prior to model
integration. In developing the merged
algorithm, some modifications were made to the original automaton
rules to more realistically mimic tumor progression. The merged
simulation tool is summarized as follows:

\begin{enumerate}

 \item \textbf{Automaton cell generation}: A Voronoi tessellation of random
 points generated using the nonequilibrium procedure of random sequential
 addition of hard disks determines the underlying lattice for our
 algorithm \cite{kansal00a, torquato02}. Here a uniform
 density lattice is used instead of the lattice with variable density.
 Each automaton cell created via this
 procedure represents a cluster of a very small number of biological cells ($\sim 10$).

 \item \textbf{Define confining boundary}:  Each automaton cell is divided
 into one of two regimes: nonmalignant cells within the confining boundary and
 nonmalignant cells outside of the boundary.

 \item \textbf{Healthy microvascular network}: The blood vessel network which
 supplies the cells in the tissue region of interest with oxygen and nutrients
 is generated using the random analog of the Krogh cylinder model detailed
 in Ref. \cite{gevertz06}. One aspect of the merger involved limiting
 blood vessel development to the subset of space in which tumor growth occurs.

 \item \textbf{Initialize tumor}: Designate a chosen nonmalignant cell inside
 the growth-permitting environment as a proliferative cancer cell.

 \item \textbf{Tumor growth algorithm}: Time is then discretized into units
 that represent one real day.  At each time step:

  \begin{enumerate}
   \item \textit{Solve PDEs}: A previously-developed system of partial
   differential equations \cite{gevertz06} is numerically solved one day
   forward in time.  The quantities that govern vasculature evolution, and
   hence are included in the equations, are concentrations of VEGF ($v$),
   unoccupied VEGFR-2 receptors ($r_{v0}$), the VEGFR-2 receptor occupied with
   VEGF ($r_{v}$), Ang-1 ($a_1$), Ang-2 ($a_2$), the unoccupied angiopoietin
   receptor Tie-2 ($r_{a0}$),  the Tie-2 receptor occupied with Ang-1 ($r_{a1}$)
   and the Tie-2 receptor occupied with Ang-2 ($r_{a2}$).  The parameters in
   these equations include diffusion coefficients of protein $x$ ($D_x$),
   production rates $b_x$ and $\overline{b}_x$, carrying capacities $K_x$,
   association and dissociation rates ($k_{y}$ and $k_{-y}$) and decay rates
   $\mu_x$.  Any term with a subscript $i$ denotes an indicator function; for
   example, $p_i$ is a proliferative cell indicator function.  It equals 1 if a
   proliferative cell is present in a particular region of space, and it equals
   0 otherwise.  Likewise, $h_i$ is the hypoxic cell indicator function, $n_i$
   is necrotic cell indicator function and $e_i$ is the endothelial cell
   indicator function.  The equations solved at each step of the algorithm are:

\begin{equation}
 \frac{\partial v}{\partial t} = D_v\Delta v + b_vh_i\big(h-v^2/K_v\big) - k_0vr_{v0} + k_{-0}r_v - \mu_{v}v
\end{equation}
\begin{equation}
 \frac{\partial a_1}{\partial t} = b_{a1}e_i(p_i+h_i+n_i)\big(e_0-a_1^2/K_a\big) - k_1a_1r_{a0} + k_{-1}r_{a1} - \mu_{a1}a_1
\end{equation}
\begin{equation}
\begin{array}{c}
\!\!\!\!\!\!\!\!\!\!\! \displaystyle{ \frac{\partial a_2}{\partial
t}} = D_{a2}\Delta a_2
+ b_{a2}e_i(p_i+h_i+n_i)\big(e_0-a_2^2/K_a\big) \\
 \quad\quad+ \overline{b}_{a2}h_i\big(h-a_2^2/K_a\big) 
 - k_2a_2r_{a0} + k_{-2}r_{a2} - \mu_{a2}a_2
\end{array}
\end{equation}
\begin{equation}
 \frac{\partial r_{v0}}{\partial t} = -k_0vr_{v0} + k_{-0}r_v
\end{equation}
\begin{equation}
 \frac{\partial r_{a0}}{\partial t} = -k_1a_1r_{a0} + k_{-1}r_{a1} - k_2a_2r_{a0} + k_{-2}r_{a2}
\end{equation}
\begin{equation}
 \frac{\partial r_v}{\partial t} = k_0vr_{v0} - k_{-0}r_v
\end{equation}
\begin{equation}
 \frac{\partial r_{a1}}{\partial t} = k_1a_1r_{a0} - k_{-1}r_{a1}
\end{equation}
\begin{equation}
 \frac{\partial r_{a2}}{\partial t} = k_2a_2r_{a0} - k_{-2}r_{a2}
\end{equation}

\noindent In these equations, $h(x,y,t)$ represents the
concentration of hypoxic cells and $e_0$ represents the
endothelial cell concentration per blood vessel.  The system of
differential equations contains 21 parameters, 13 of which were
taken from experimental data.  Parameters were unable to be found
in the literature were estimated.  For more details on the
parameter values, as well as information on the initial and
boundary conditions and the numerical solver, the reader is
referred to Ref \cite{gevertz06}.


   \item \textit{Vessel Evolution}: Check whether each vessel meets the
   requirements for regression or growth.  Vessels with a concentration of bound
   Ang-2 six times greater than that of bound Ang-1 regress \cite{maisonpierre97},
   provided that the concentration of bound VEGF is below its critical value.
   Vessel tips with a sufficient amount of bound VEGF sprout along the VEGF gradient.

   \item \textit{Nonmalignant Cells}: Healthy cells undergo apoptosis if vessel
   regression causes its oxygen concentration to drop below a critical threshold
   (more particularly, if the distance of a healthy cell from a blood
   vessel exceeds the assumed diffusion length of oxygen, 250 $\mu$m~).
   Further, nonmalignant cells do not divide in the model.  While
   nonmalignant cell division occurs in some organs, a hallmark of neoplastic
   growth is that tumor cells replicate significantly faster than the corresponding
   normal cells. Hence, we work under the simplifying assumption that nonmalignant
   division rates are so small compared to neoplastic division rates that they
   become relatively unimportant in the time scales being considered. In the
   cases where this assumption does not hold, nonmalignant cellular division would
   have to be incorporated into the model.

   \item \textit{Inert Cells}: Tumorous necrotic cells are inert. This
   assumption is certainly valid for the tumor type that motivated this modeling
   work, glioblastoma multiforme. In the case of glioblastoma, the presence of
   necrosis is an important diagnostic feature and, in fact, negatively
   correlates with patient prognosis. 

   \item \textit{Hypoxic Cells}: A hypoxic cell turns proliferative if its
   oxygen level exceeds a specified threshold \cite{gevertz06} and turns necrotic
   if the cell has survived under sustained hypoxia for a specified number of days.
   In the original algorithms, the transition from hypoxia to necrosis was based on
   an oxygen concentration threshold.  However, given that cells (both tumorous
   and nonmalignant alike) have been shown to have a limited lifespan under
   sustained hypoxic conditions, a temporal switch more accurately
   describes the hypoxic to necrotic transition.  Thus, a novel aspect
   of the merged algorithm is a temporal hypoxic to necrotic transition.  It has
   been measured that human tumor cells remain viable in hypoxic regions of a
   variety of xenografts for 4-10 days \cite{gevertz06}.  In our simulations, we
   will use the upper-end of this measurement and assume that tumor cells can
   survive under sustained hypoxia for 10 days.

   \item \textit{Proliferative Cells}: A proliferative cell turns hypoxic if its
   oxygen level drops below a specified threshold.  However, if the oxygen level
   is sufficiently high, the cell attempts to divide into the space of a viable
   nonmalignant cell in the growth-permitting region. The probability of
   division $p_{div}$ is given by:

     \begin{equation}
      p_{div} = p_0(1-r/L_{max})
     \end{equation}

   where $p_0$ is the base probability of division, $r$ is the distance of the
   dividing cell from the geometric center of the tumor and $L_{max}$ is the
   distance between the closest boundary cell in the direction of tumor growth
   and the tumor's center.  In the original implementations of the algorithm,
   $p_0$ was fixed to be 0.192, giving a cell-doubling time of
   $\ln(2)/\ln(1+p_0)\approx$ 4 days.  In the merged algorithm proposed here,
   we wanted to account for fact that tumor cells with a higher oxygen
   concentration likely have a larger probability of dividing than those with a
   lower oxygen concentration.  For this reason, we have modified the algorithm
   so that $p_0$ depends on the distance to the closest blood vessel $d_{vessel}$
   (which is proportional to the oxygen concentration at a given cell site). The
   average value of $p_0$ was fixed to be 0.192, and we have specified that $p_0$
   takes on a minimum value $p_{min}$ of 0.1 and a maximum value $p_{max}$ of 0.284.
   This means that a proliferative cell in the model can have a cell doubling
   time anywhere in the range of three to seven days.  The formula used
   to determine $p_0$ is
    \begin{equation}
     p_0 = \frac{\a p_{min}-p_{max}}{\a D_{O2}}d_{vessel} + p_{max},
    \end{equation}
   where $D_{O2}$ is the diffusion length of oxygen, taken to be
   250 $\mu$m \cite{gevertz08, gevertz06}. Both $p_{min}$ and $p_{max}$ depend on the
   average probability of division.  If this average probability changes, so
   does $p_{min}$ and $p_{max}$.
   \item \textit{Tumor Center and Area}: After each cell has evolved,
   recalculate the geometric center and area of the tumor.

  \end{enumerate}

\end{enumerate}

The readers are referred to Ref. \cite{gevertz09} for more
details, including how  cell-level phenotypic heterogeneity is also
considered in a similar fashion to the manner
done in Refs \cite{kansal00b} and \cite{schmitz02}.

\subsection{Simulating Heterogeneous Tumor Growth}


The 3D cancer simulation tool described here was
employed to study tumor growth in a confined environment: a two-dimensional
representation of the cranium in space as a function of time \cite{gevertz09}. 
The cranium is idealized as an elliptical growth-permitting
environment with two growth-prohibiting circular obstacles
representing the ventricular cavities. Tumor growth is initiated
in between a ventricular cavity and the cranium wall. In this
setting, we find that the early-time characteristics of the tumor
and the vasculature are not significantly different than those
observed when radial symmetry is imposed on tumor growth. In
particular, after 45 days of growth (Figure 16(a)),
vessels associated with the radially symmetric tumor begin to
regress and hypoxia results in the tumor center. Twenty days later
(Figure 16(b)), a strong, disordered angiogenic response
has occurred in the still radially symmetric tumor. Over the next
50 days of growth (Figure 16(c) and (d)), the
disorganized angiogenic blood vessel network continues to
vascularized the growing tumor, but the tumor's shape begins to
deviate from that of a circle due to the presence of the confining
boundary. The patterns of vascularization observed are consistent
with the patterns observed in the original vascular model
\cite{gevertz06}, suggesting that the merged algorithm maintains
the functionality of the original vascular algorithm.

\centerline{Figure 16: Tumor growing in a 2D
representation of the cranium.}

However, if the results of this simulation are compared with those
of the environmentally-constrained algorithm without the explicit
incorporation of the vasculature \cite{gevertz08},
we find that the merged model responds to the environmental
constraints in a way that is more physically intuitive. In the
original environmentally-constrained algorithm \cite{gevertz08},
the tumor responds quickly and drastically to the
confining boundary and ventricular cavities. This occurs because
the original evolution rules not only determine the probability of
division based on the distance to the boundary, but also determine
the state of a cell based on a measure of its distance to the
boundary. In the merged model which explicitly incorporates the
vasculature, the state of each cell depends on the blood vessel
network, and only the probability of division directly depends on
the boundary.  For this reason, the merged algorithm exhibits an
emergent property in that it grows tumors that respond more
gradually and naturally to environmental constraints than does the
algorithm without the vasculature.

The tumor growth in a two-dimensional irregular region of space that truly allows the
neoplasm to adapt its shape as it grows in time (i.e., a 3D model) was also investigated
by Gevertz {\textit et al} \cite{gevertz08}; see also Gevertz and Torquato \cite{gevertz09}.
As with the two-dimensional representation of the cranium in space, an emergent property
of the merged algorithm in which that a more subtle and natural
response to the effects of physical confinement is found occurs.
The studies taking into account mutations responsible for
phenotypic heterogeneity have been carried out by Gevertz and
Torquato \cite{gevertz09}, to which the readers are referred for more
details. We note that all the results presented in this section
need to be validated experimentally.

\section{Analysis of the Invasion Network: Minimal Spanning Trees}

It is well known that cancer cells can break off the main tumor
mass and invade healthy tissue.  For many cancers, this
process can eventually result in metastases to other organs.
Tumor-cell invasion is a hallmark of
glioblastomas, as individual tumor cells have been observed to
spread diffusely over long distances and can migrate into regions
of the brain essential for the survival of the patient \cite{holland00}.
In certain cases, the invading tumor cells
form branched chains (see Figure 3), i.e., tree structures \cite{kansal01}.
The brain offers these invading cells a variety of pathways
they can invade along (such as blood vessel and white fiber
tracts), which may be interpreted as the edges of an underlying
graph with the various ``resistances" values along these pathways playing
the role of edge weights. The underlying physics behind the
formation of the observed patterns are only beginning to be
understood.

\centerline{Figure 17: Examples of weighted graph and the
resulting minimal spanning tree.}

The competition between local and global driving forces
is a crucial factor is determining the structural organization
 in a wide variety of
naturally occurring branched networks \cite{Me94,Ro97,We97}. As an
attempt toward a model of the invasive network emanating from a
solid tumor, Kansal and Torquato \cite{kansal01} investigated the
impact of a global minimization criterion versus a local one on
the structure of spanning trees. Spanning trees are defined as a
loopless, connected set of edges that connect all of the nodes in
the underlying graph (see Figure 17). In particular,
these authors considered the generalized minimal spanning tree
(GMST) and generalized invasive spanning tree (GIST), because they
generally offer extremes of global (GMST) and local (GIST)
criteria. Both GMST and GIST are defined on graphs in which the
nodes are partitioned into groups and each edge has an assigned
weight. GMST is refined (relative to that of a spanning tree) such
that the requirement that every node of the graph is included in
the tree is replaced by the inclusion of at least one node from
each group with the additional requirement that the total weight
of tree is minimized \cite{dror00}. GIST can be constructed by
growing a connected cluster of edges by ``invading'' the remaining
the edge with the minimal weight at its boundary with the
requirement of the inclusion of at least one node from each group
in the final tree \cite{kansal01}.

Kansal and Torquato \cite{kansal01} have developed efficient algorithms to
generate both GMST and GIST structures, as well as a method to
convert GIST structure incrementally into a more globally
optimized GMST-like structure (see Figure 18). The readers are referred to the
original paper for more algorithmic details. These methods allow
various structural features to be observed as a function of the
degree to which either criterion is imposed and the intermediate
structures can then serve as benchmarks for comparison when a real
image is analyzed.

\centerline{Figure 18: Examples of GMST and GIST.}

We note that a general procedure by which information extracted
from a single, fixed network structure can be utilized to
understand the physical processes which guided the formation of
that structure is highly desirable in understanding the invasion
network of tumor cells, since the temporal development of such a
network is extremely difficult to observe. To this end, Kansal and
Torquato \cite{kansal01} examined a variety of structural characterizations
and found that the occupied edge density (i.e., the fraction of
edges in the graph that are included in the tree) and the
tortuosity of the arcs in the trees (i.e., the average of the
ratio of the path length between two arbitrary nodes in the tree
and the Euclidean distance between them) correlate well with the
degree to which an intermediate structure resembles the GMST or
GIST. Since both characterizations are straightforward to
determine from an image (e.g., only the information of the tree is
required for tortuosity and additional information of underlying
graph is needed for occupied edge density), they are potentially
useful tools in the analysis of the formation of invasion network
structures. Once the distribution of the invasive cells in the
brain is understood, a cellular automaton simulation tool for
glioblastoma that is useful in a clinical setting could be
developed. This of course would apply more generally to
invasion networks around other solid tumors.

\section{Conclusions and Future Work}

In this paper, we have reviewed the work that we have performed to 
attempt to develop
an Ising model of cancer. We began by describing a
minimalist 4D cellular automaton  model of cancer in which
healthy cells transition between states (proliferative, hypoxic,
and necrotic) according to simple local rules and their present
states, which can viewed as a stripped-down Ising model of cancer
\cite{kansal00a, kansal00b, schmitz02}.
Using four proliferation parameters, this model was
shown to reflect the growth dynamics of a clinically
untreated tumor very well \cite{kansal00a}. This was followed by
discussion of an extension of the model to study the effect on the
tumor dynamics and geometry of a mutated subpopulation \cite{kansal00b}
and how tumor growth is affected by chemotherapeutic
treatment, including induced resistance, with additional three
treatment parameters \cite{schmitz02}. An improved CA model
that explicitly accounts for  angiogenesis \cite{gevertz06} as
well as the heterogeneous and confined environment in which a
tumor grows \cite{gevertz08} were discussed. A general cancer
simulation tool that merges, adapts and improves all of the
aforementioned mechanism into a single CA model
was also presented and applied to simulate the growth
the GBM in a vascularized confined cranium \cite{gevertz09}.
Finally, we touched on how one might characterize the invasive
network organization (local versus global organization) around a solid tumor using spanning trees \cite{kansal01}.
However, we must move well beyond the improved CA model
as well as other computational models of cancer 
in order to make real progress on controlling this
dreaded set of diseases.

\subsection{The Obvious but Necessary}

Formulating theoretical and computational tools that can be
utilized clinically to predict neoplastic progression and
propose individualized optimal treatment strategies to control
cancer growth is the holy grail of tumor modeling.  Although the
development of our most comprehensive cellular automaton model is
potentially a useful step towards the long-term goal of an Ising model for cancer, numerous complex
mechanisms involved in tumor growth and their interactions needs to be
identified and understood in
order to truly achieve this goal.

For example, an effective Ising model of cancer
must incorporate  molecular-level
information via a better understanding of the cellular origin of
the tumor. Such information might become available if
 imaging techniques for spatial statistics of
cell/molecular heterogeneity can be developed. This would
enable an improved understanding of invading  cancer cells:
cell motility, cell-cell communication and phenotypes
of invading cells.  Such knowledge is crucial
in order to predict the effects of treatment and
tumor recurrence. The incorporation  of
stem cells, oncogenes and tumor suppressor genes
in computational models would aid in our understanding
of tumor progression.

In addition, we must quantitatively characterize the biological
(host) environment (i.e., a heterogeneous material/medium) in
which cancer evolves, including both the microstructure and the
associated physical properties. For example, a better knowledge of
diffusion and transport of nutrients, drugs, etc. would
significantly improve the accuracy of the model simulating the
effects of vasculature evolution and treatment. Similarly, cell
mechanics and mechanical stresses must be understood.
In such cases, imaging of the
biological environment over a wide spectrum of length and time
scales will be crucial.


\centerline{Figure 19: A cartoon of a two-phase medium.}

It is important to emphasize  that the theory
heterogeneous media is a huge field within the physical
sciences that can be brought to bear to better understand the host
heterogeneous microenvironment of cancer and metastases (see Figure 19).
 For example, there exist powerful and
sophisticated theoretical/computational techniques to
characterize the microstructure of heterogeneous materials
and predict their physical properties \cite{torquato02}.
Specifically, the details of
complex microstructures are described in terms of various statistical
descriptors (different types of correlation functions), which in turn
determine the physical properties of the heterogeneous materials \cite{torquato02}.
In particular, the effective properties that have been predicted
include the diffusion coefficient \cite{to99}, reaction rates \cite{Le89a,to91},
elastic/viscoelastic moduli \cite{quin95, to97},
thermal conductivity \cite{Ki91a}, thermal expansion coefficient \cite{gibi97},
fluid permeability \cite{To90}, and
electrical conductivity \cite{To85f,sen89}. Accurate characterizations of these
properties of the host environment and tumor mass are essential
in order to significantly improve models for tumor growth and invasion.
For example, a knowledge of the elastic properties enables one
to better model the effects of physical confinement and the
mechanical response of solid tumor; while the diffusion coefficient
and fluid permeability are crucial to model transport of nutrients and proteins,
delivery of drugs and even the migration of cancer cells.
These techniques have been used to propose a
novel biologically constrained three-phase model of the brain microstructure \cite{Ge08a}.

Given such information, the CA model can be modified accordingly
to take into account the available cell/molecular details of the
tumor mass, its invasion network and the host heterogeneity (e.g.,
the capillary vasculature and adaptive physical confinement).
Real-time tumor growth and treatment simulations can be carried
out to generate data of clinical utility. For example, instead of
only producing data which qualitatively reflects the general
effects of tumor treatment and resistance, one could use the
model to make reliable prognosis and to optimize individual
treatment strategy.

It would be fascinating to see if a more refined Ising model for cancer 
predicted a ``phase transition'' phenomenon, which would 
be in keeping with the behavior of the standard Ising model 
for spin systems. For example, it is not hard to imagine that 
part of the tumorigenesis process involves a ``phase transition'' 
between pre-malignant cells and malignant cells.

\subsection{Not So Obvious: Optimization and Cancer}

We also note that variational principles \cite{torquato02,to04} and
optimization techniques \cite{sig97, hyun02, To04b,to10} have been
fruitfully applied to design structures with optimal properties.
Can optimization techniques be applied to understand and control cancer? 
Although optimization methods have begun to be employed
for such purposes, there full potential has yet to be 
realized. For tumor treatment, for example,
optimization techniques could be employed to
design chemotherapy/radiation strategies depending on tumor size, genomic
information and the heterogeneous environment as well as the optimal durations of
treatment and rest periods. Given sufficient patient-specific information,
optimized treatment strategies can be designed for individual patients.
A variety of optimization techniques could be brought to bear here,
including simulated annealing methods, and  linear and nonlinear
programming techniques.

\centerline{Figure 20: Minimal surface structure}

We have  developed an optimization methodology that provides a means of optimally
designing multifunctional composite microstructures 
\cite{hyun02, to10}. We have shown how the competition between two
different performance demands (thermal versus electrical
behaviors or electrical versus mechanical behaviors) results in
unexpected microstructures, namely, minimal
surfaces \cite{hyun02,To04b} (see Figure 20), which
also appear to optimal for fluid transport \cite{Ju05}
as well as diffusion-controlled reactions \cite{Ge09a}. This work
suggests that it may be fruitful to explore the development of
cancer, which not only involves competition but cooperation, from a rigorous 
multifunctional optimization viewpoint.
 Cancer processes involve a competition
between the primary clone, sub-clones, healthy tissue, immune
system, etc. as well as a cooperation between different cells types (e.g., stroma cells
and cancer cells) in a heterogeneous environment. This competition/cooperation can
be translated into an optimization problem in space and time.
Adaptation of this multifunctional optimization approach 
to cancer modeling could provide an alternative to game-theory approaches
to understanding cancer \cite{Di09}.

\subsection{The Far Out}

\centerline{Figure 21: Diamond and disordered ground
state.}

Even more challenging and intriguing questions can be asked: Can we exploit the
unique properties of normal stem cells \cite{reya01} to control
cancer (e.g., to deliver therapy to tumors or to have them compete
with the tumor)? Can we use inverse optimization methods to design
``hypothetical'' cancers or stem cells with particular the
cell-cell interactions to yield targeted behaviors and then make
them in the lab? These ``inverse'' problems are motivated by their
analog in statistical mechanics \cite{rech05, rech07, bat09,
to09}. In statistical mechanics, the ``forward problem'' is one in
which a Hamiltonian (interaction potential) for a many-body system
is specified and then structure of the system  and its
thermodynamics  are predicted. By contrast,  the ``inverse''
problem of statistical mechanics  seeks to find the ``optimal''
interaction potential that leads spontaneously to a novel ``targeted"
structure (or behavior). We have discovered optimal interaction potentials
that yield unusual or counterintuitive targeted ground (zero-temperature) states,
e.g., low-coordinated diamond crystal \cite{rech07} and disordered states
\cite{bat09} with only isotropic pair potentials (see Figure
21). Ground states are those many-particle configurations
that arise as a result of slowly cooling a liquid
to absolute zero temperature. The aforementioned obtained targeted ground states
are so unusual because much of our experience involves ground states
that are highly-coordinated crystal structures \cite{to09}.
An extremely challenging and fascinating question is
whether we can devise inverse optimization techniques  to control
cancer?

It is clear that theoretical methods based in the physical and
mathematical sciences offer many different fruitful ways to
contribute to tumor research. However, for this approach to be
successful, intensive interactions with cell biologists,
oncologists, radiologists, clinicians, physicists, chemists,
engineers, and applied mathematicians are
essential. Such an interdisciplinary approach appears to be
necessary in order to control this deadly disease. This
could be achieved most effectively if we could
have an analog of the ``Manhattan Project'' in which
there was a single facility with such an interdisciplinary
team of scientists dedicated to this supreme achievement.

\noindent{\bf Acknowledgments}
\smallskip

\noindent{The author thanks Yang Jiao and Jana Gevertz for
very useful discussions and their critical reading of
this manuscript. The research described was supported by Award Number U54CA143803 from the National Cancer Institute.  The content is solely
the responsibility of the authors and does not necessarily represent the official views of the National Cancer Institute or
the National Institutes of Health.}

\bibliographystyle{unsrt}
\newpage

\section*{References}

\newpage

\section*{Glossary}
\begin{itemize}

 \item \textit{Neoplasm}: A neoplasm is a synonym for a tumor.

 \item \textit{Glioma}: A collection of tumors arising from the glial cells
 or their precursors in the central nervous system.

 \item \textit{Cellular automaton}: A spatially and temporally discrete model
 that consists of a grid of cells, with each cell being in one of a number of
 predefined states.  The state of a cell at a given point in time depends on
 the state of itself and its neighbors at the previous discrete time point.
 Transitions between states are determined by a set of local rules.

 \item \textit{Ising model}: The Ising model is an idealized statistical-mechanical model of ferromagnetism
that is based on simple local-interaction rules, but nonetheless
lead to basic insights and features of real magnets, such as phase transitions with a critical point.

 \item \textit{Voronoi cell}: Given a set of points, the Voronoi cell is the
 cell that is formed about an arbitrary point in the set by finding the
 region of space closer to that point than any other point in the
 system (Torquato, 2002).

 \item \textit{Delaunay triangulation}: Given a Voronoi graph (a set of
 Voronoi cells), the Delaunay graph is its dual that results from joining all
 pairs of sites that share a Voronoi face.  If this graph consists of only
 simplices, the graph is called a Delaunay triangulation (Torquato, 2002).

 \item \textit{Quiescent}: A cell is considered quiescent if
 it is in the G0 phase of the cell cycle and is not actively dividing.

 \item \textit{Necrotic}: A cell is considered necrotic if it has died due
 to injury or disease, such as abnormally low oxygen levels.

\end{itemize}

\newpage

\section*{Figure Legends}

     \begin{figure}[ht]
     \label{fig1}
   \centering
   \includegraphics[width=0.45\textwidth]{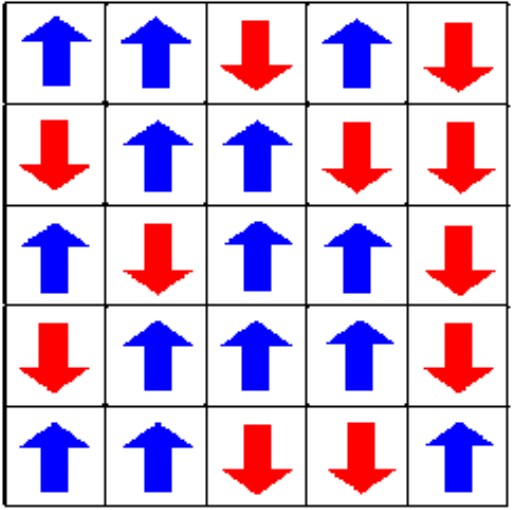}
   \caption{A schematic plot of the Ising model for an idealized ferromagnet.
The model consists of spins that can be in one of two states (up or down)
arranged in this case on a square lattice. In its simplest rendition,
 each spin interacts only with its nearest neighbors. Such simple
local interaction rules can result in rich collective behavior
depending on the temperature of the system.}
   \end{figure}

\newpage
  \begin{figure}[ht]
  \label{fig2}
  \centering
  \includegraphics[width=0.35\textwidth]{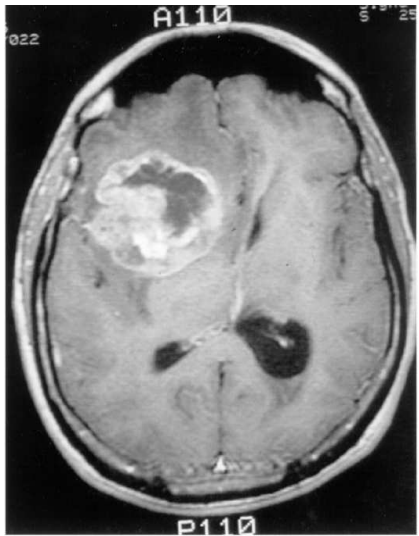}
  \caption{A $T_1$-contrast enhanced brain MRI-scan
   showing a right frontal GBM tumor, as adapted from Ref. \cite{kansal00a}. Perifocal hypointensity is
   caused by significant edema formation. The hyper-intense, white
   region (ring-enhancement) reflects an area of extensive blood-brain/tumor
   barrier leakage. Since this regional neovascular setting provides
   tumor cells with sufficient nutrition it contains the highly metabolizing, e.g. dividing, tumor cells
   Therefore, this area corresponds to the
   outermost concentric shell of highly proliferating neoplastic cells in
   our model (see Figure 5).}
  \end{figure}

\newpage
     \begin{figure}[ht]
   \centering
   \includegraphics[width=0.45\textwidth]{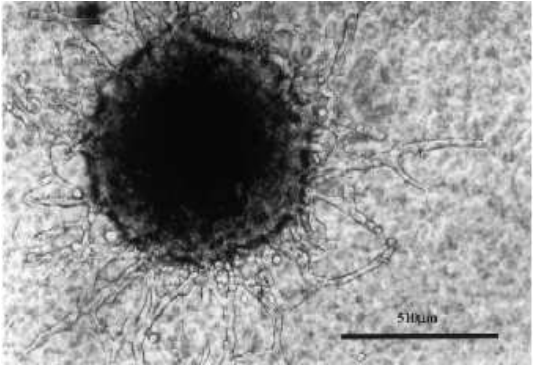}
   \caption{MTS-gel assay showing a central spheroid with multiple
   ``chain''-like invasion pathways leading towards the boundary
   (magnification: x 200), as adapted from Ref. \cite{kansal00a}. }
   \label{fig3}
   \end{figure}

\newpage
  \begin{figure}[ht]
  \label{fig4}
   \centering
 $\begin{array}{c@{\hspace{2cm}}c}
   \includegraphics[width=0.25\textwidth]{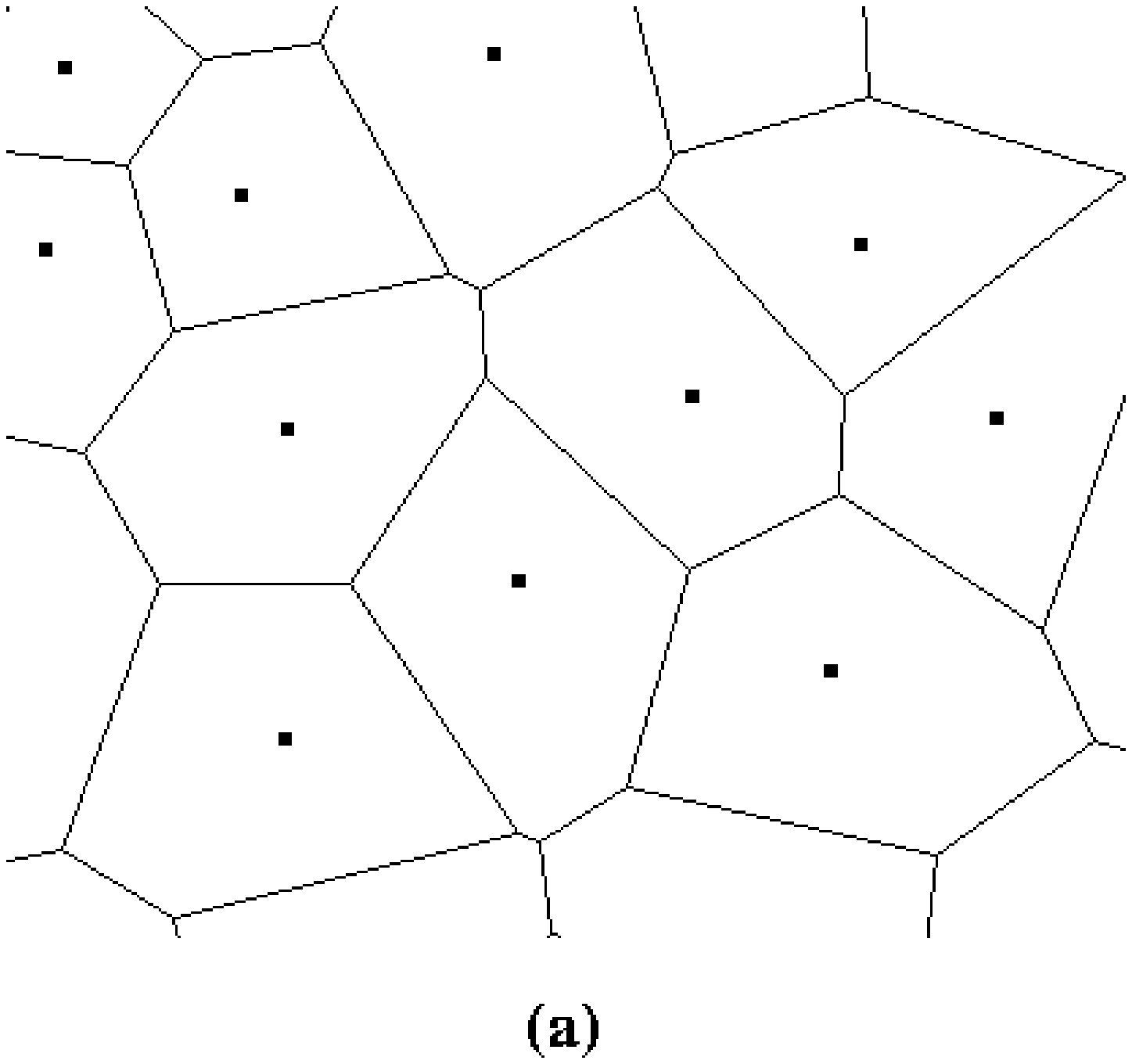} &
 \includegraphics[width=0.25\textwidth]{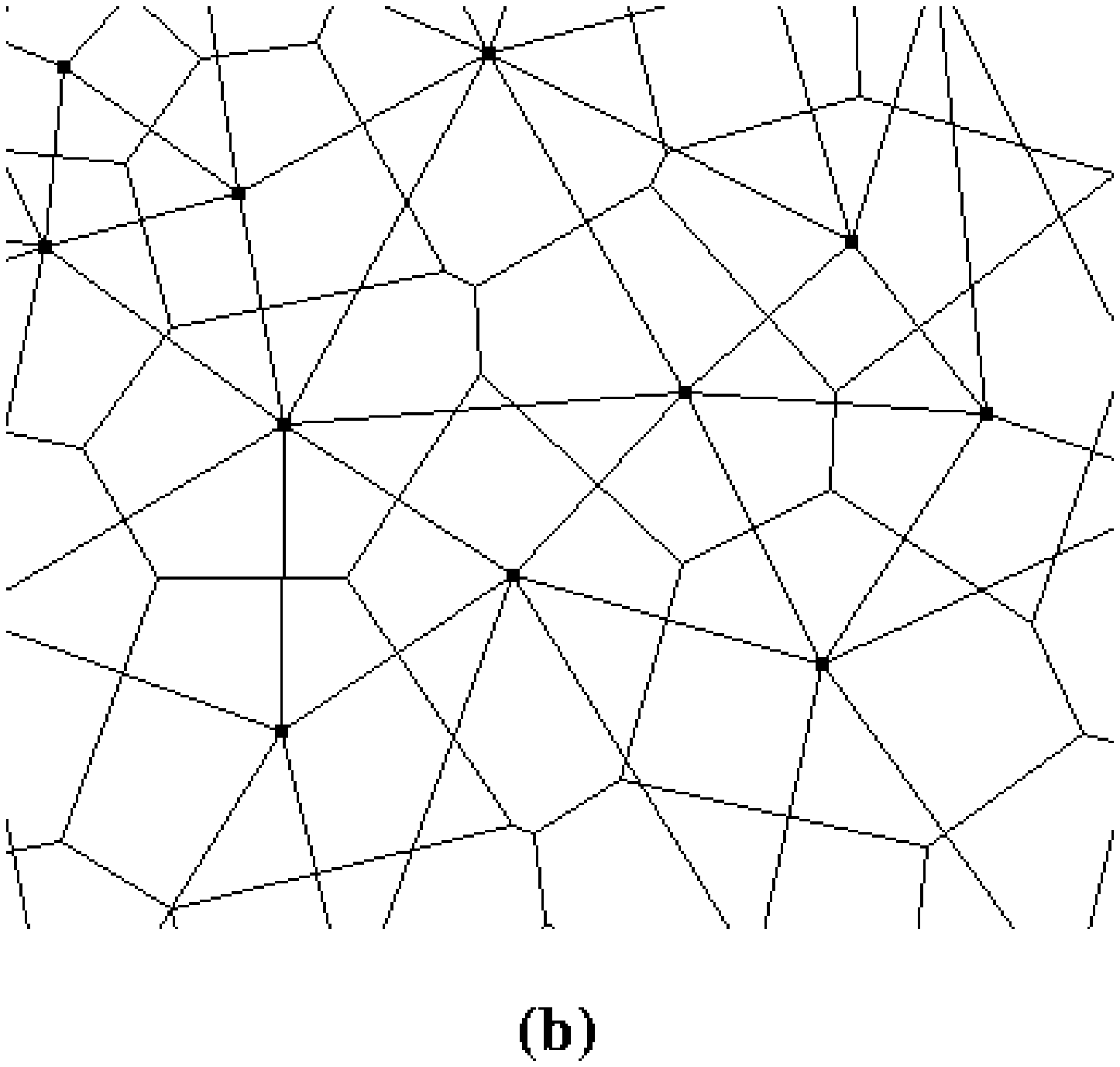} \\
\includegraphics[width=0.25\textwidth]{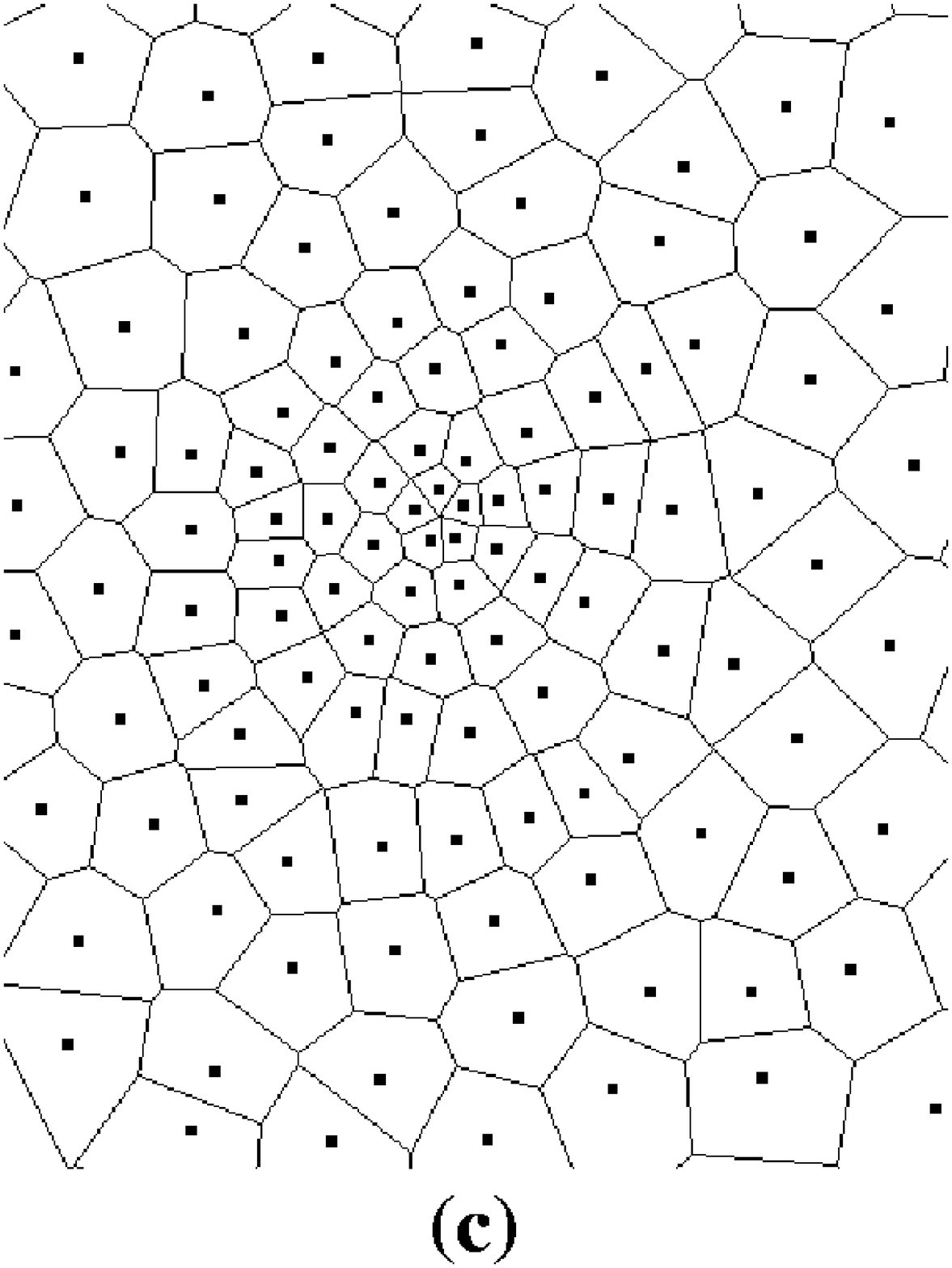} &
 \includegraphics[width=0.25\textwidth]{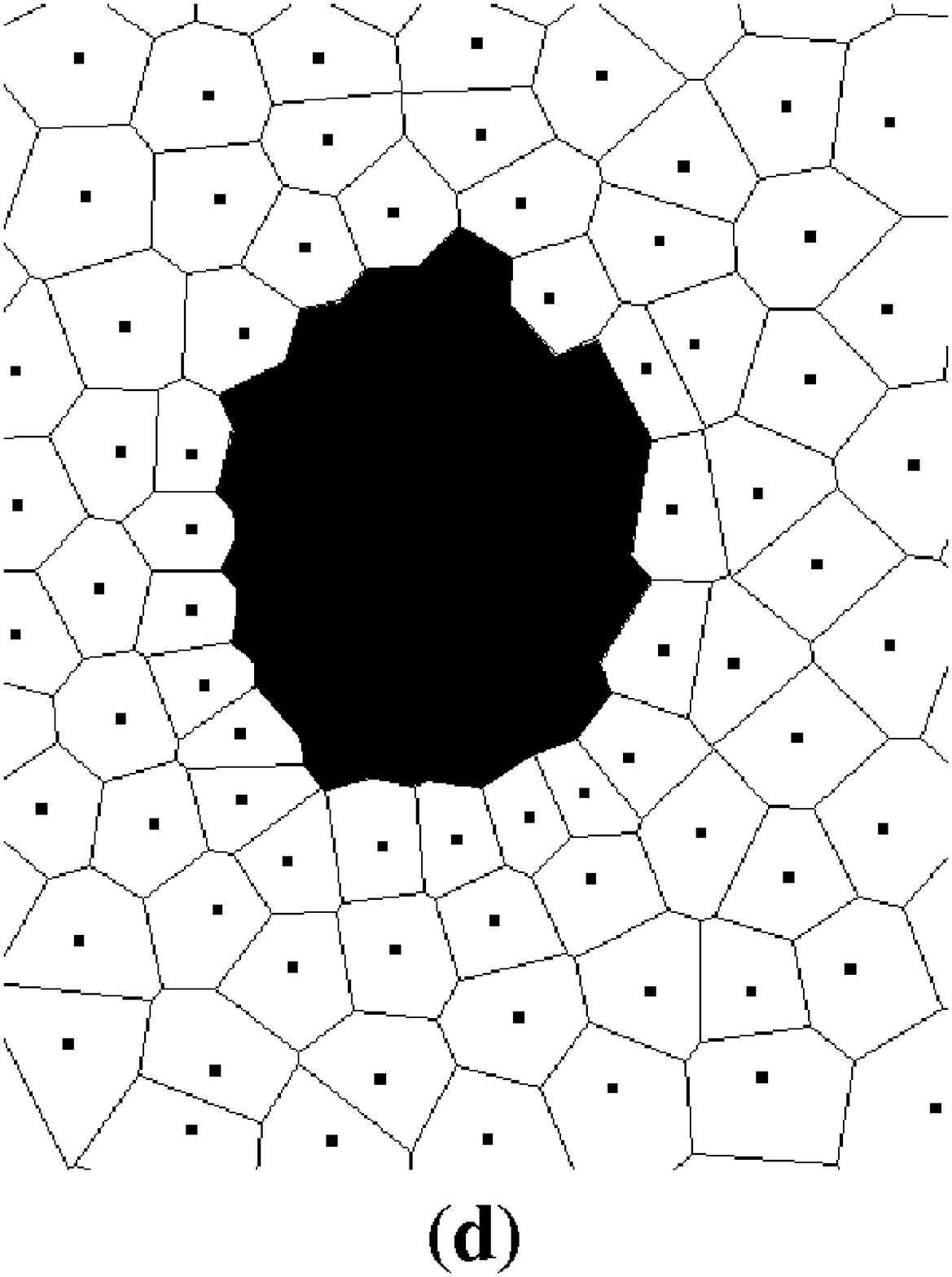} \\
\end{array}$
   \caption{Two-dimensional space tiled into Voronoi cells,
as adapted from Ref. \cite{kansal00a}. 
   Points represent sites and lines denote boundaries between cells.  Figures
   (a) and (b) depict a very small section of a lattice.  (a) shows the
   Voronoi cells, while (b) shows both the Voronoi cells, along with the
   Delaunay tessellation. Figures (c) and (d) show a more representative section
   of the lattice, with the variable density of sites evident.
   Panel (c) shows the entire lattice section, (d) shows the same section
   with the darkened cells representing a tumor.}
   \end{figure}

\newpage
\begin{figure}[ht]
\label{fig5}
   \centering
   \includegraphics[width=0.35\textwidth]{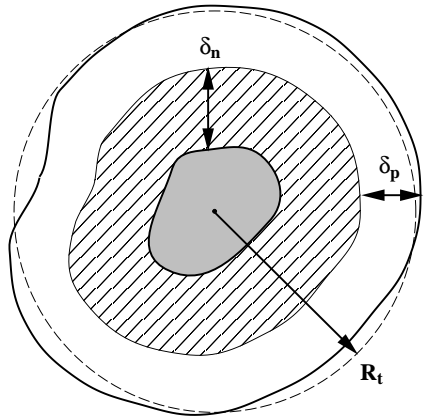}
   \caption{A cross-section of an idealized solid tumor, 
as adapted from Ref. \cite{kansal00a}.  The inner gray region is composed
   of necrotic tissue. The cross-hatched layer is composed of
   living, quiescent cells (non-proliferative).  It has a thickness
   $\delta_n$. The outer shell, with thickness $\delta_p$, is composed of
   proliferative cells. Both length scales $\delta_n$ and $\delta_p$ are
determined by nutritional needs of the cells via diffusional transport.}
   \end{figure}

\newpage
\begin{figure}[ht]
  \label{fig6}
   \centering
 \includegraphics[width=0.9\textwidth]{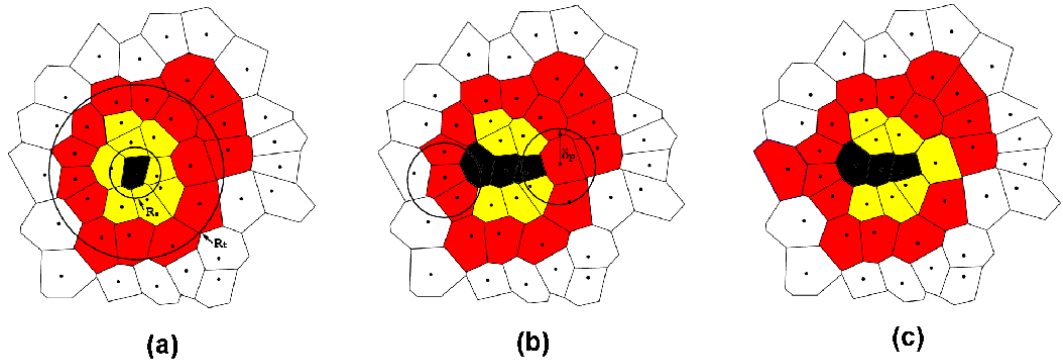}
   \caption{An illustration of the minimalist proliferation algorithm through
a cross-section of the solid tumor \cite{kansal00a}. (a) A tumor contains
   necrotic (black), non-proliferative (yellow or light gray) and
   proliferative cells (red or dark gray). The average overall tumor radius $R_t$
   and the necrotic region radius $R_n$ are shown. (b) Two non-proliferative
   cells that more than $\delta_n$ away from the tumor edge are turned into
   necrotic and two proliferative cells are selected with probability $p_d$ to
   check for division. If there are non-tumorous cells within a distance $\delta_d$
   from the selected proliferative cell, it will divide; otherwise, it will
   turn into a non-proliferative cell. (c) One of the selected
   proliferative cell divides and the other turns into a non-proliferative cell.}
   \end{figure}

\newpage
\begin{figure}[ht]
\label{fig7}
   \centering
   \includegraphics[width=0.65\textwidth]{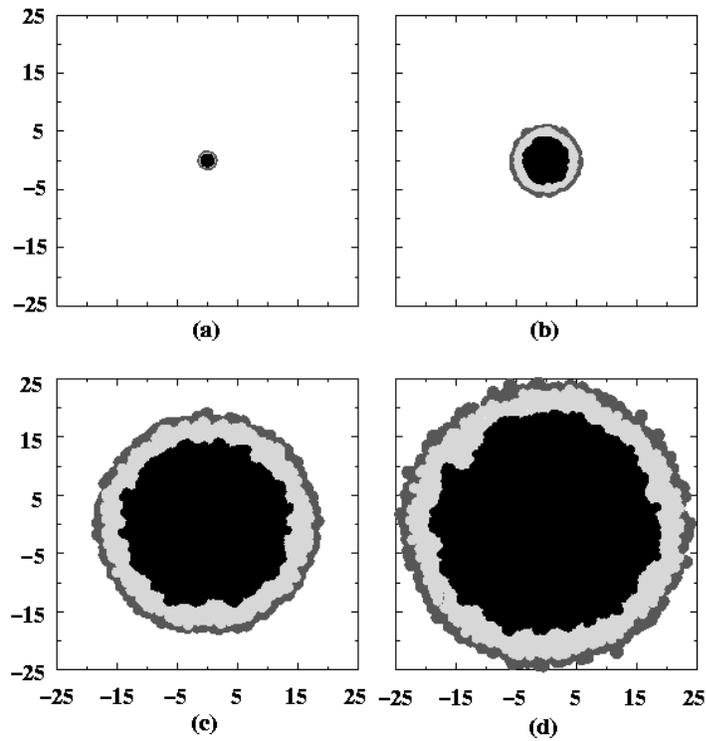}
   \caption{The development of the central section of a tumor as a function of
   time,  as adapted from Ref. \cite{kansal00a}. Correspond to: (a) the initial  tumor spheroid stage, (b)
   time to first detectable
   lesion, (c) time at diagnosis and (d) time at death. The dark gray outer region is
   comprised of proliferating cells, the light gray is non-proliferative cells
   and the black is necrotic cells. The length scales are given in millimeters.}
   \end{figure}

\newpage
\begin{figure}[ht]
\label{fig8}
   \centering
   \includegraphics[width=0.35\textwidth]{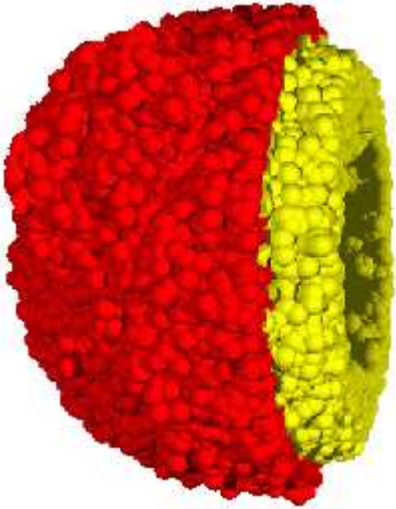}
   \caption{A cut-away view a simulated tumor generated from the
minimalist CA algorithm \cite{kansal00a}. The inner necrotic core
   is not depicted in this view. The yellow (light gray) region is
   comprised of nonproliferative cells and the red (dark gray) shell
   depicts the proliferative cells.}
   \end{figure}

\newpage
\begin{figure}[ht]
\label{fig9}
   \centering
   \includegraphics[width=0.5\textwidth]{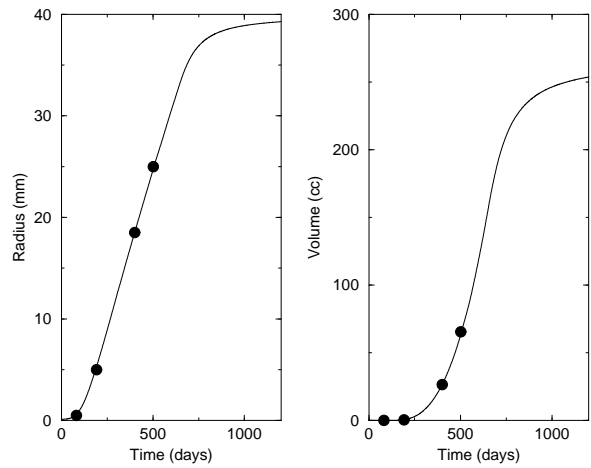}
   \caption{Plots of the radius and volume of the tumor versus
   time, as adapted from Ref. \cite{kansal00a}.  The lines correspond to simulation predictions,
   using the first parameter set given in the text.
   The plotted points reflect the test case derived from
   the medical literature. A quantitative comparison of the simulation
   with the test case is given in Table 2.}
   \end{figure}

\newpage
\begin{figure}[ht]
\label{fig10}
   \centering
    $\begin{array}{c@{\hspace{2cm}}c}
   \includegraphics[width=0.45\textwidth]{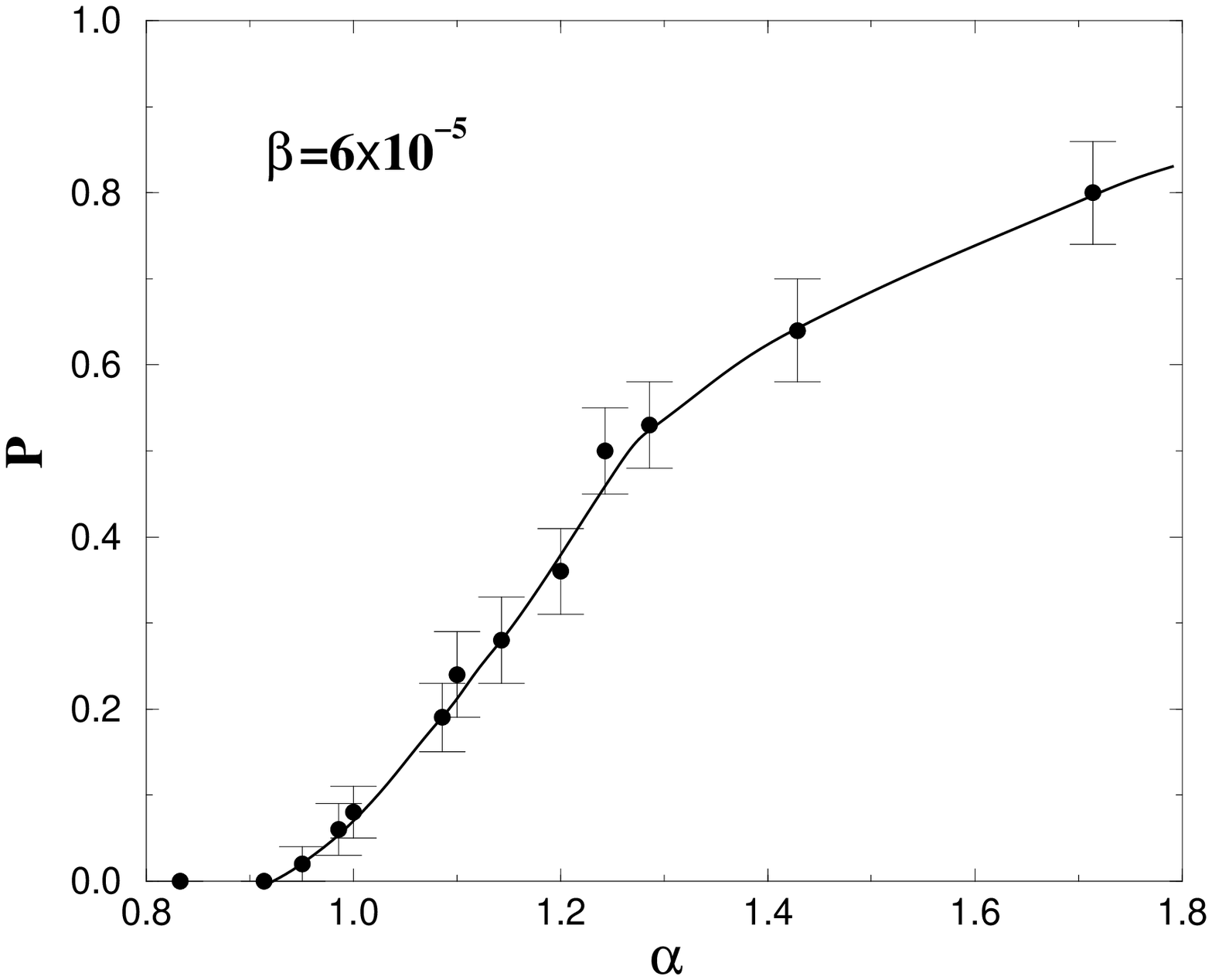} &
 \includegraphics[width=0.25\textwidth]{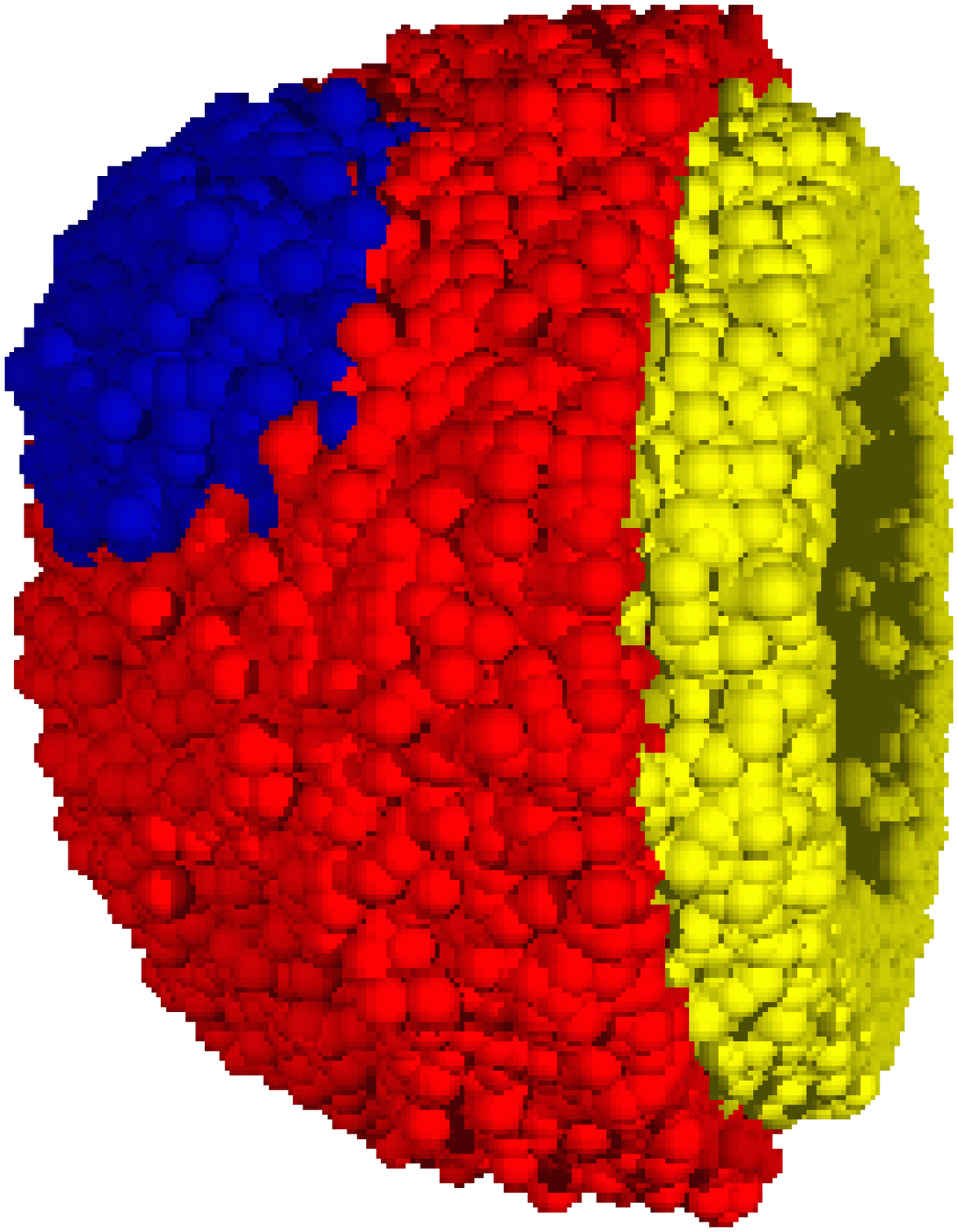} \\
\mbox{{\bf (a)}} & \mbox{{\bf (b)}}\\
\end{array}$
   \caption{(a) A plot of probability of emergence $P$ versus
   the degree of mutation $\alpha$, i.e., growth advantage ($\alpha >1$)
or disadvantage ($\alpha <1$), as adapted from Ref. \cite{kansal00b}. 
   The error bars indicate confidence intervals defined by one standard
   deviation from the mean. Each data point represents the
   average of roughly 100 simulated tumors.  The line is drawn as a guide
   for the eye. (b) A cut-away view a simulated tumor with a mutated population. The inner necrotic core
   is not depicted in this view. The yellow (light gray) region is
   comprised of nonproliferative cells and the red (dark gray) shell
   depicts the proliferative cells of the primary strain and the blue (darker gray)
   shows the proliferative cells of the secondary strain.}
\end{figure}

\newpage
\begin{figure}[ht]
\label{fig11} \centering
 \includegraphics[width=0.6\textwidth]{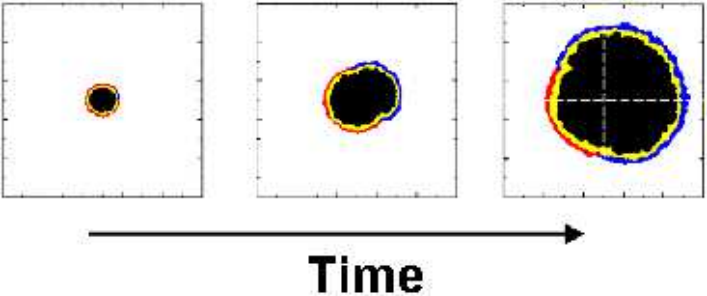}
\caption{The effects of the emergence of a subpopulation on the
tumor geometry, as adapted from Ref. \cite{kansal00b}. The center of mass of the tumor is significantly
shifted as the growth of the subpopulation, }
\end{figure}

\newpage
\begin{figure}[ht]
\label{fig12} \centering
 \includegraphics[width=0.55\textwidth]{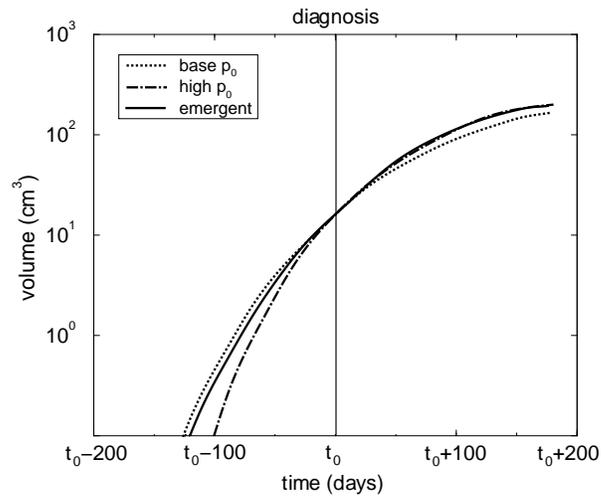}
\caption{Volume of a simulated
tumor with an emergent subpopulation in time, as adapted from Ref. \cite{kansal00b}.  Volumes of tumors
composed entirely of the primary strain and the secondary strain
are also shown and labeled ``base $p_0$'' and ``high $p_0$,''
respectively.  Each tumor is set to have the same volume at some
``diagnosis'' time $t_0$.  Note that the emerging tumor's dynamics
initially follow the base case, but later follow the highly
aggressive case.}
\end{figure}

\newpage
\begin{figure}[ht]
\label{fig13} \centering
 \includegraphics[width=0.6\textwidth]{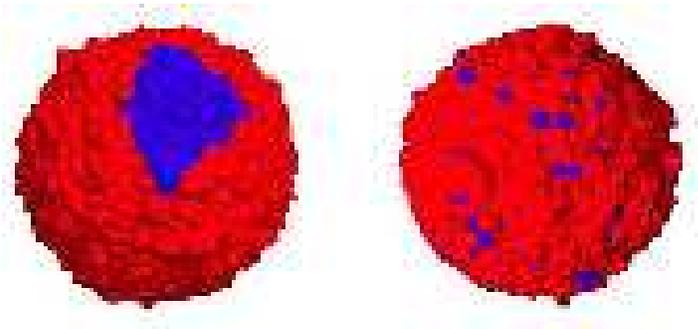}\\
\caption{Initial images of two-strain tumors, as adapted from Ref. \cite{schmitz02}. 
The resistant subpolulation is localized (left
panel) and scattered (right panel). The blue cells
of each tumor belong to the resistant subpopulation, while
the blues ones belong to the sensitive subpopoluation.}
\end{figure}

\newpage
\begin{figure}[ht]
\label{fig14} \centering
 \includegraphics[width=0.6\textwidth]{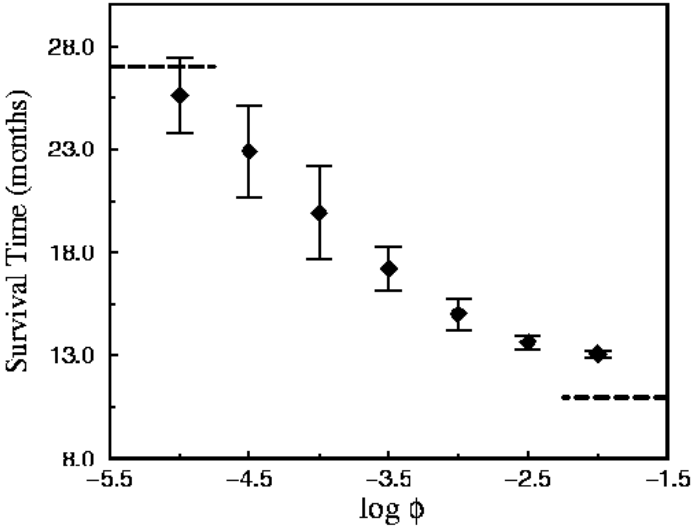}\\
\caption{Survival times
associated with continuously mutating tumors,  as adapted from Ref. \cite{schmitz02}. 
This figure depicts data of the mean survival time (with error
bars) as a function of $\phi$, the expected fraction
of tumor cells that mutate at each instance of treatment.}
\end{figure}

\newpage
\begin{figure}[ht]
\label{fig15} \centering
 \includegraphics[width=0.65\textwidth]{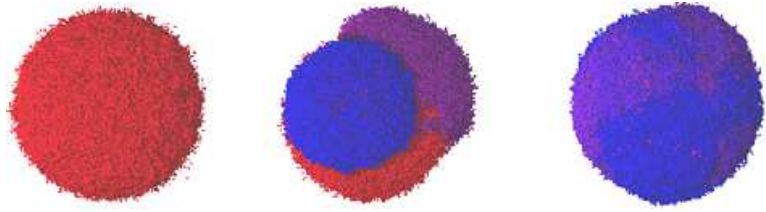}
\caption{Images of
continuously mutating tumors,  as adapted from Ref. \cite{schmitz02}. Shown are
   representative images of tumors with small $\phi$ (left panel),
   intermediate $\phi$ (middle panel) and large $\phi$ (right panel).
   The distinct clonal sub-populations in each tumor are
   represented with a different color, ranging from red (highest $\gamma$-values)
   to violet (lowest $\gamma$-values). All tumors here are fully-developed.}
\end{figure}

\newpage
\begin{figure}[ht]
\label{fig16} \centering
 \includegraphics[width=0.65\textwidth]{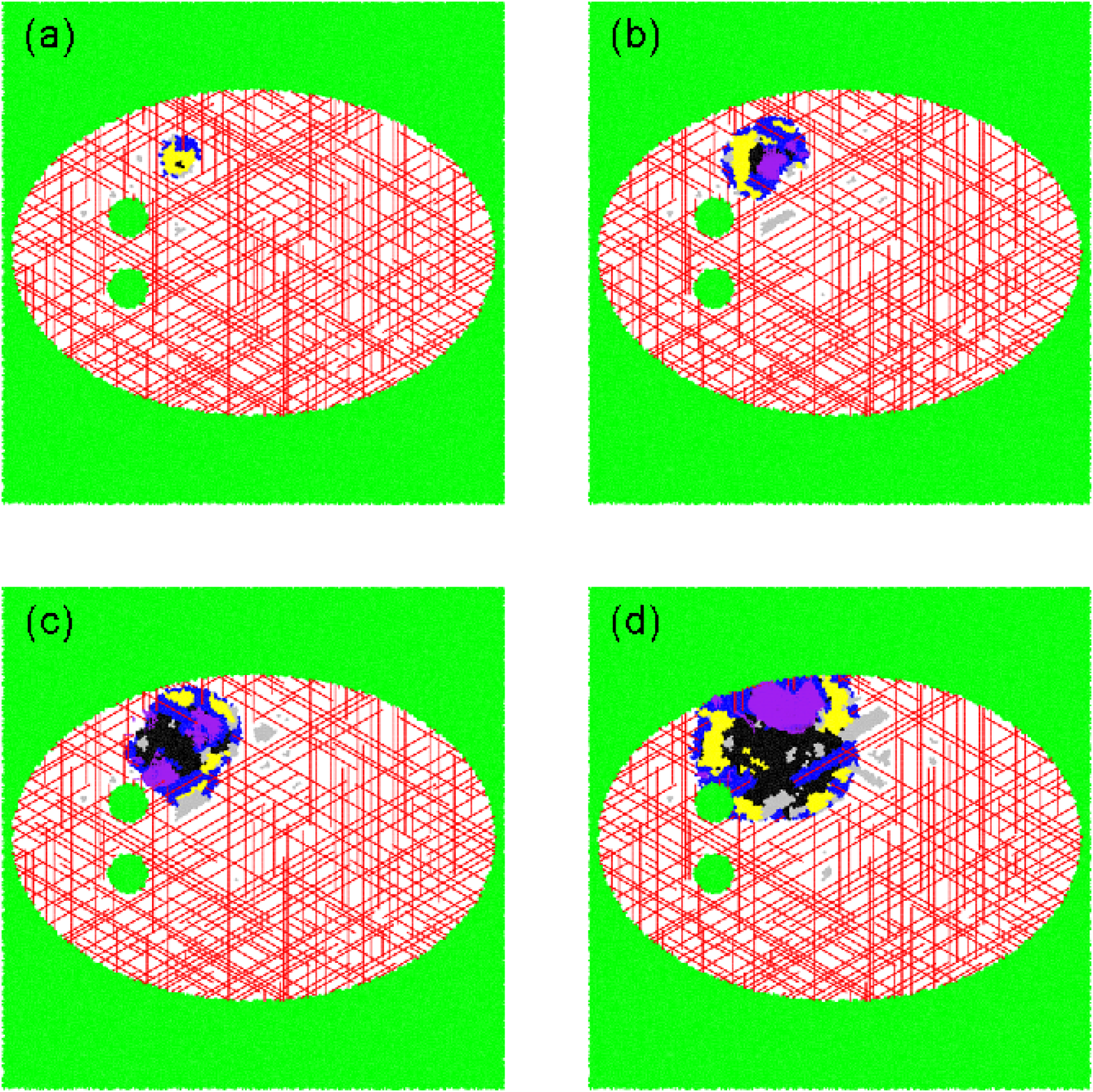}
\caption{The temporal
development of a tumor growing in
   a  two-dimensional representation of the cranium in space, as adapted from  Ref. \cite{gevertz09}. (a) After 40 days, the dimensionless area
   is 0.0049 units$^2$, with 30\% of the cells being proliferative, 66.4\% being
   hypoxic and 3.6\% being necrotic.  (b) After 65 days, the dimensionless area is
   0.0195 units$^2$, with 51.2\% of the cells being proliferative, 33.0\% being
   hypoxic and 15.8\% being necrotic.  (c) After 85 days, the dimensionless area is
   0.0362 units$^2$, with 48.2\% of the cells being proliferative, 16.8\% being
   hypoxic and 35.0\% being necrotic. (d) After 115 days, the dimensionless area is
   0.0716 units$^2$, with 45.1\% of the cells being proliferative, 18.6\% being
   hypoxic and 36.3\% being necrotic.  The deep blue outer region (darkest of the
   grays in black and white) is comprised of proliferative cells, the yellow region
   (lightest of the grays in black and white) consists of hypoxic cells and the black
   center contains necrotic cells. Green cells (intermediate gray shade in black and
   white) are apoptotic.  The white speckled region of space represents locations
   in which the tumor cannot grow.  The lines represent blood vessels.  If viewing
   the image in color, red vessels were part of the original tissue vasculature,
   and the purple vessels grew via angiogenesis.}
\end{figure}

\newpage
\begin{figure}[ht]
\label{fig17}
   \centering
 $\begin{array}{c@{\hspace{0.5cm}}c@{\hspace{0.5cm}}c}
   \includegraphics[width=0.25\textwidth]{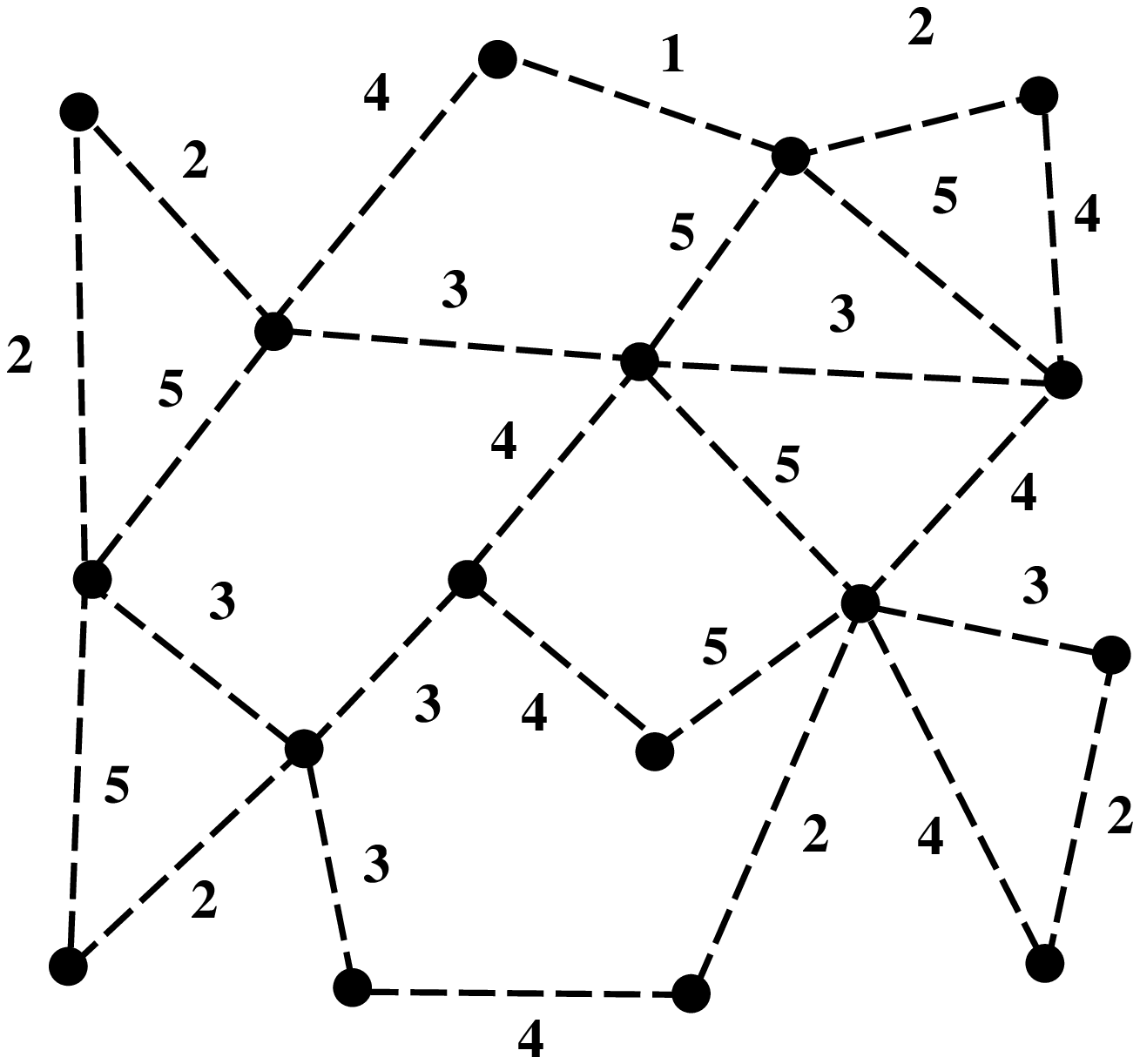} &
 \includegraphics[width=0.25\textwidth]{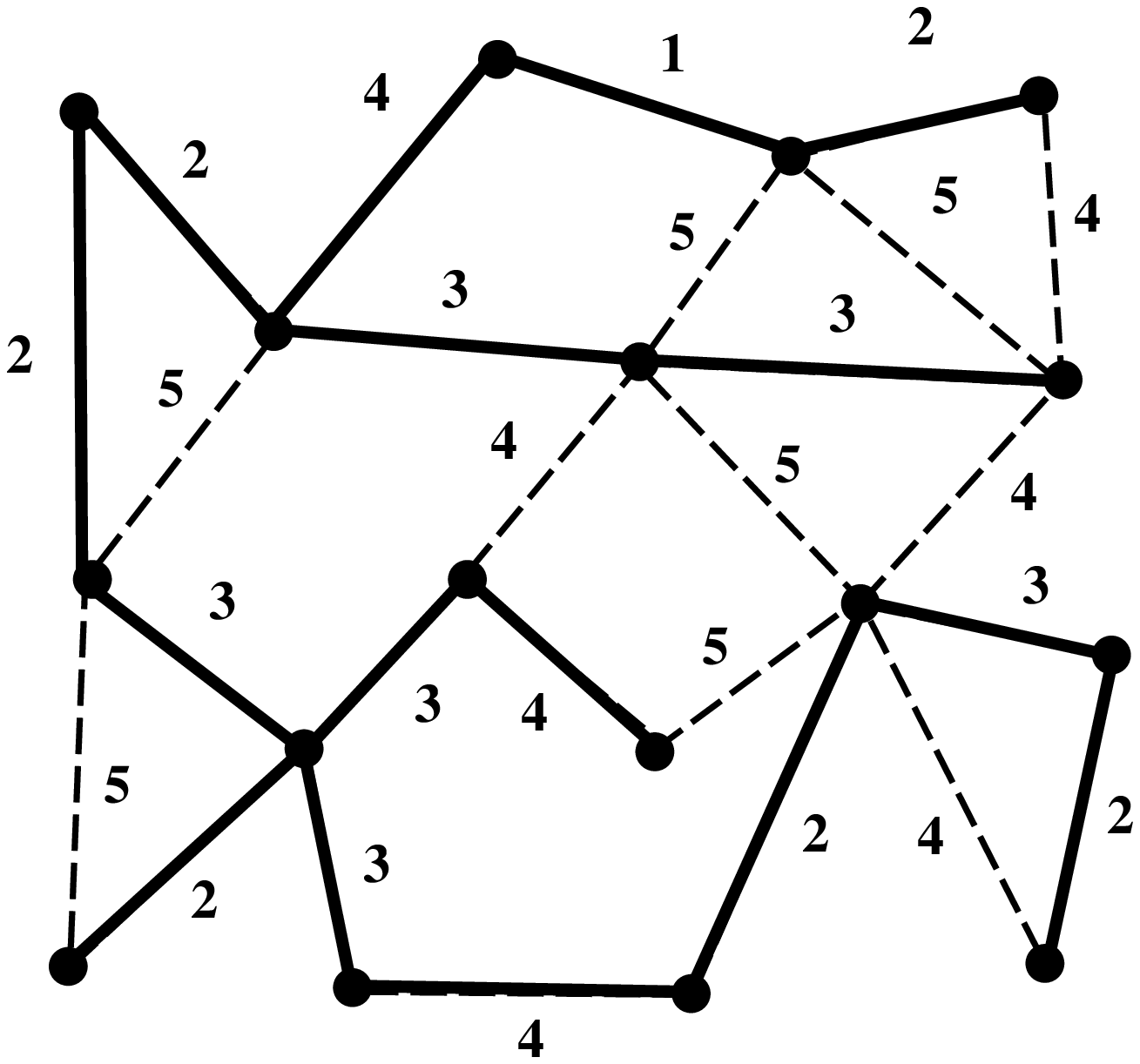} &
 \includegraphics[width=0.25\textwidth]{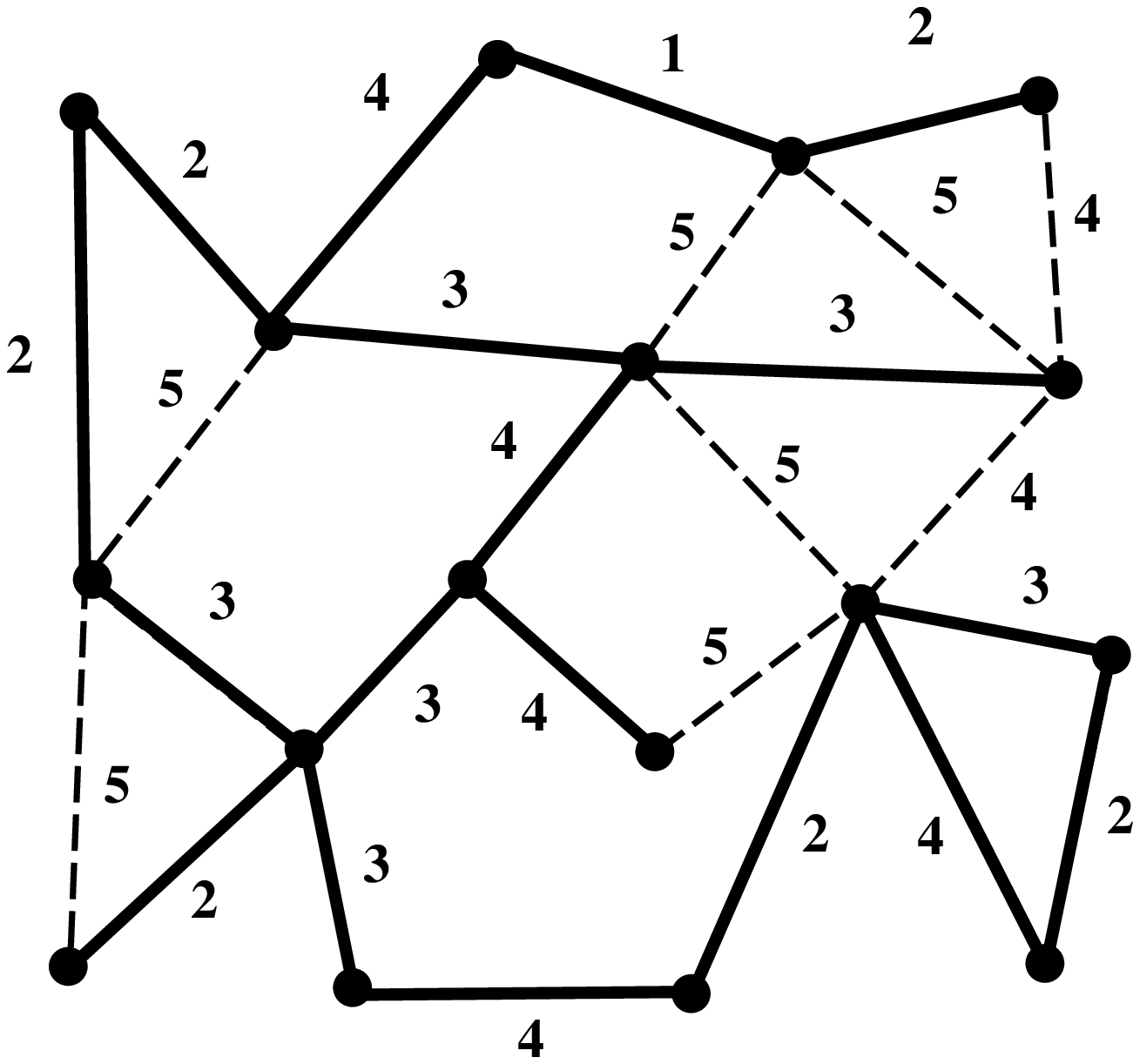} \\
\end{array}$
   \caption{Example of a weighted graph and the
   resulting minimal spanning tree, as adapted from Ref. \cite{kansal01}.  (a)~Shows all of the edges and nodes in a graph,
   with the weight of each edge indicated next to the edge.
   Graph edges are depicted by broken lines.
   (b)~Shows the minimal spanning tree for this graph, which is the
   set of edges that connects every node in the graph in the tree with
   the lowest total weight. Edges included in the tree are shown as
   solid lines, while edges not included remain broken lines. The total
   weight of the tree in (b) is 40, and the occupied edge density
   (number of edges included in the tree divided by total number of edges
   in the graph) is $15/25 = 0.6$. (c)~Shows the invasion percolation network
   for the same graph.  Note that the invasion percolation network may have
   loops and in this case there are two closed loops.  If loop formation is
   prevented (resulting in the highest weight edge in any loop remaining
   unoccupied) the result is the acyclic invasion percolation network.
   As can be readily seen by comparing figures (b) and (c) the acyclic
   invasion percolation network is identical to the minimal spanning tree.}
   \end{figure}

\newpage
\begin{figure}[ht]
\label{fig18}
   \centering
 $\begin{array}{c@{\hspace{1.5cm}}c}
 \includegraphics[width=0.35\textwidth, clip=]{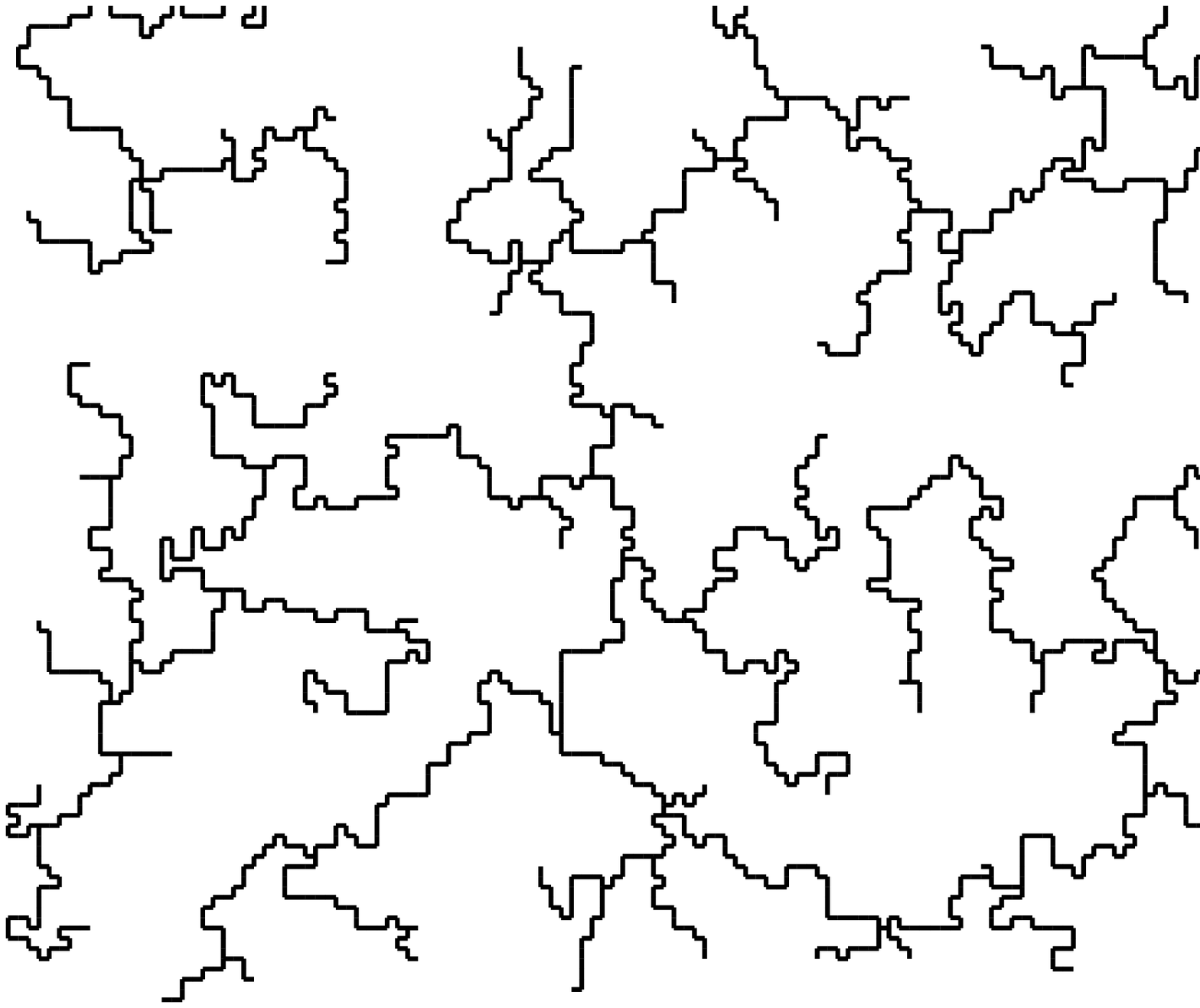} &
 \includegraphics[width=0.35\textwidth, clip=]{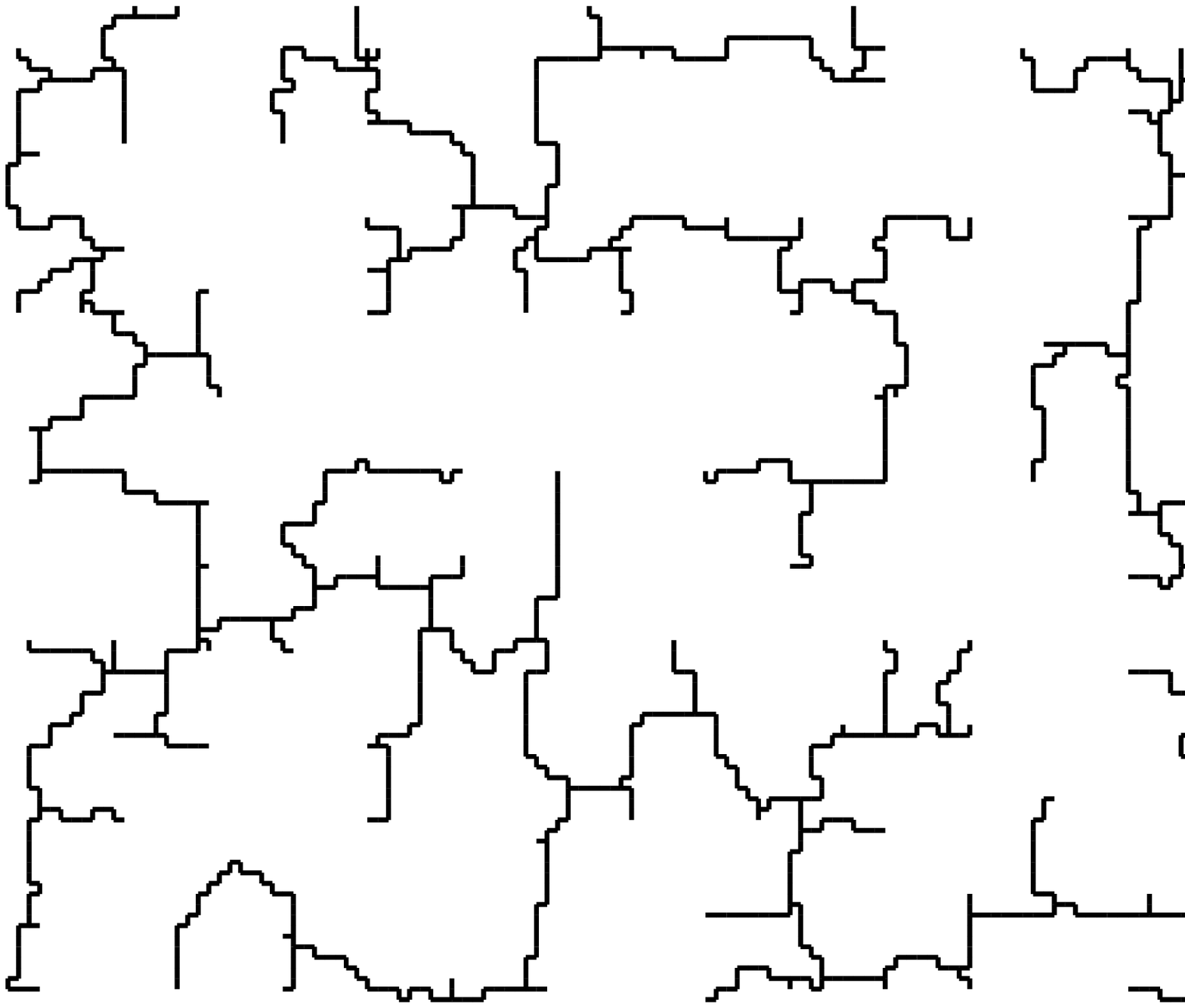} \\
\end{array}$
   \caption{Examples of (a) backbone of generalized invasive spanning tree (GIST) (b)
   generalized minimal spanning tree (GMST), as adapted from Ref. \cite{kansal01}.}
   \end{figure}

\newpage
\begin{figure}[ht]
\label{fig19} \centering
\includegraphics[width=0.85\textwidth]{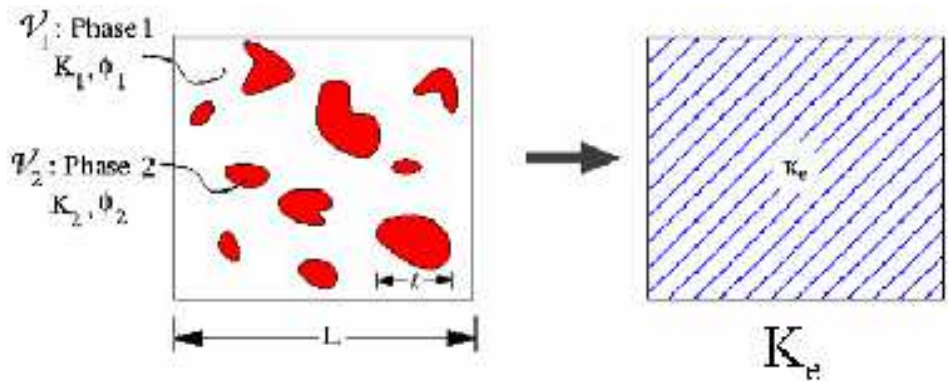}
\caption{A schematic illustration of heterogeneous materials, as adapted
from Ref. \cite{torquato02}.
   Left panel: A two-phase heterogeneous material with properties ${\bf K}_1$ and ${\bf K}_2$
and volume fractions $\phi_1$ and $\phi_2$. The quantity $K_i$ represents
any general physical property of phase $i$ (e.g., diffusion coefficient, electrical or thermal
conductivity, elastic moduli, viscosity, and magnetic permeability). 
The material phases can either be solid, liquid or gas depending
on the specific context. Here $L$ and $\ell$
represent the macroscopic and microscopic length scales, respectively.
    Right panel: When $L$ is much bigger than $\ell$, the heterogeneous
material can be replaced by  homogeneous medium with 
    an effective property ${\bf K}_e$.}
\end{figure}

\newpage
\begin{figure}[ht]
\label{fig20}
\centering
\includegraphics[width=0.6\textwidth]{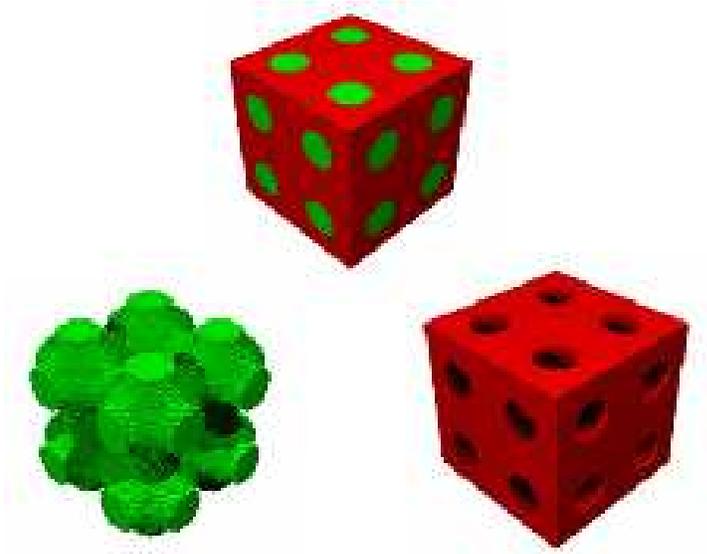}
\caption{Two-phase composite whose interface is the Schwartz P minimal surface, which
   simultaneously optimizes heat and electrical conduction \cite{hyun02}
or electrical conduction and elastic moduli \cite{To04b}. The image at the top
shows the two-phase composite. The bottom left image shows only one of the phases
(green), which can be seen to be the Schwartz P minimal surface, and the bottom right image shows only the other phase (red).}
\end{figure}

\newpage
\begin{figure}[ht]
\label{fig21}
   \centering
 $\begin{array}{c@{\hspace{1.5cm}}c}
 \includegraphics[width=0.35\textwidth, clip=]{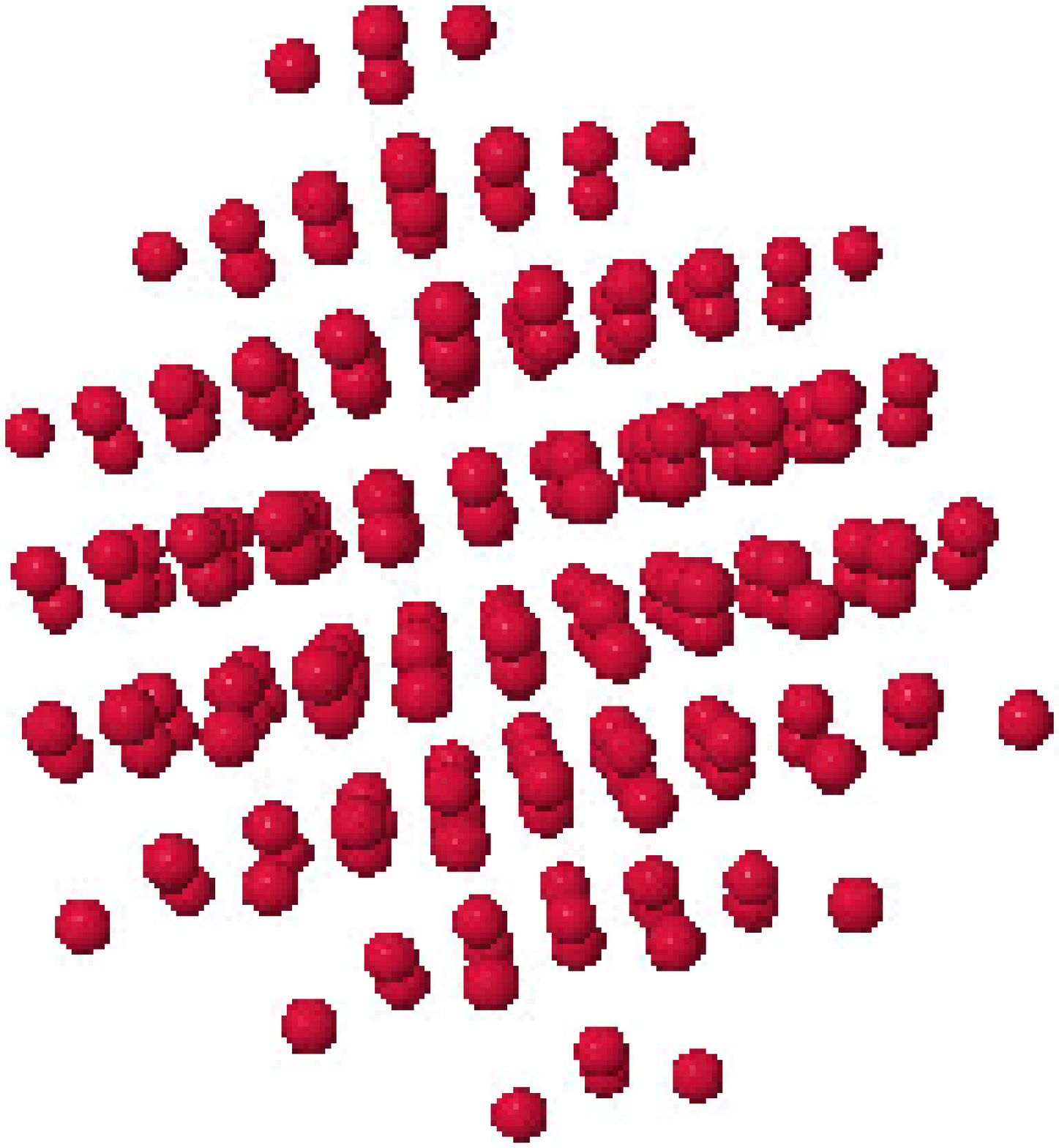} &
 \includegraphics[width=0.35\textwidth, clip=]{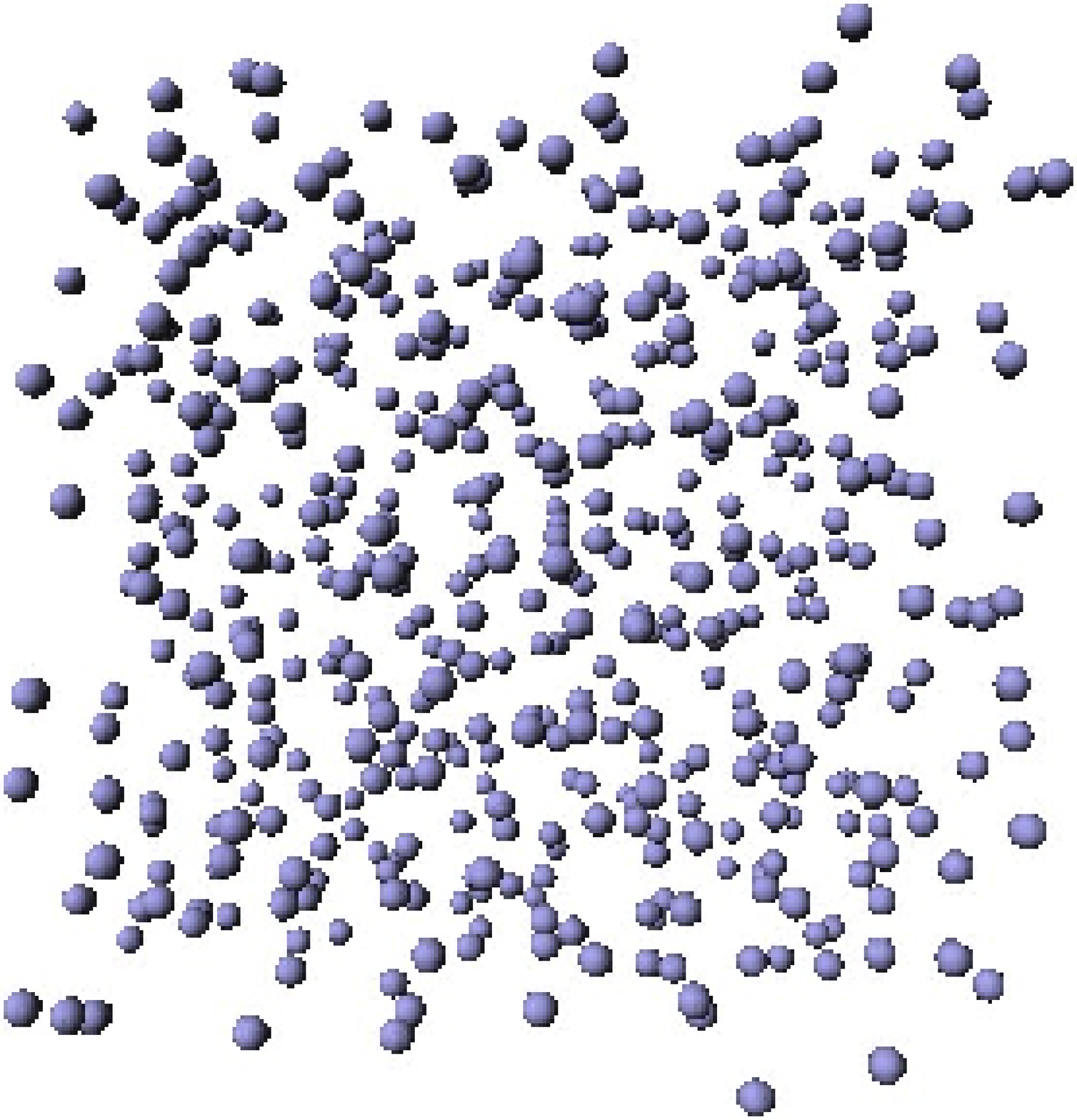} \\
\end{array}$
\caption{The ground-state structures for isotropic pair interaction
potentials
   obtained via ``inverse'' optimization
   techniques discussed in the text. Left panel: A diamond-crystal ground state \cite{rech07}. 
Right panel: A disordered ground state \cite{bat09}.}
\end{figure}








\end{document}